\documentclass[prl,superscriptaddress,preprint,longbibliography,endfloats]{revtex4-1}
\usepackage{bm,graphicx,graphics,amsmath,amssymb,bm,epsfig,color}
\usepackage{euscript,tabularx}
\usepackage{longtable}

\usepackage{hyperref}

\begin{document}

%%%%%%%%%%%%%%

\title{Nutation spectroscopy of a nanomagnet driven into \\ deeply
  nonlinear ferromagnetic resonance}

\author{Y. Li}
\email{Currently at Materials Science Division, Argonne National Laboratory, Argonne (IL), USA}
\affiliation{Service de Physique de l'\'{E}tat Condens\'{e}, CEA, CNRS, Universit\'{e} Paris-Saclay, 91191 Gif-sur-Yvette, France}

\author{V. V. Naletov}
\affiliation{Service de Physique de l'\'{E}tat Condens\'{e}, CEA, CNRS, Universit\'{e} Paris-Saclay, 91191 Gif-sur-Yvette, France}
\affiliation{Institute of Physics, Kazan Federal University, 420008 Kazan, Russian Federation}

\author{O. Klein}
\affiliation{Universit\'{e} Grenoble Alpes, CEA, CNRS, Grenoble INP, INAC-Spintec, 38000 Grenoble, France}

\author{J. L. Prieto}
\affiliation{Instituto de Sistemas Optoelectr\'onicos y
  Microtecnolog\'{\i}a (UPM), 28040 Madrid, Spain}

\author{M. Mu\~noz}
\affiliation{Instituto de Microelectr\'onica de Madrid (CNM-CSIC),
  28760 Madrid, Spain}

\author{V. Cros}
\affiliation{Unit\'{e} Mixte de Physique CNRS, Thales, Univ. Paris-Sud, Universit\'{e} Paris-Saclay, 91767 Palaiseau, France}

\author{P. Bortolotti}
\affiliation{Unit\'{e} Mixte de Physique CNRS, Thales, Univ. Paris-Sud, Universit\'{e} Paris-Saclay, 91767 Palaiseau, France}

\author{A. Anane}
\affiliation{Unit\'{e} Mixte de Physique CNRS, Thales, Univ. Paris-Sud, Universit\'{e} Paris-Saclay, 91767 Palaiseau, France}

\author{C. Serpico}
\affiliation{Dipartimento di Ingegneria Elettrica e Tecnologie dell'Informazione, Universit\`{a} Federico II, 80138 Napoli, Italy}

\author{G. de Loubens}
\email{gregoire.deloubens@cea.fr}
\affiliation{Service de Physique de l'\'{E}tat Condens\'{e}, CEA, CNRS, Universit\'{e} Paris-Saclay, 91191 Gif-sur-Yvette, France}

\date{\today}

%%%%%%%%%%%%%%

\begin{abstract}

  Strongly out-of-equilibrium regimes in magnetic nanostructures
  exhibit novel properties, linked to the nonlinear nature of
  magnetization dynamics, which are of great fundamental and practical
  interest. Here, we demonstrate that field-driven ferromagnetic
  resonance can occur with substantial spatial coherency at
  unprecedented large angle of magnetization precessions, which is
  normally prevented by the onset of spin-wave instabilities and
  magnetization turbulent dynamics. Our results show that this
  limitation can be overcome in nanomagnets, where the geometric
  confinement drastically reduces the density of spin-wave modes. The
  obtained deeply nonlinear ferromagnetic resonance regime is probed
  by a new spectroscopic technique based on the application of a
  second excitation field. This enables to resonantly drive slow
  coherent magnetization nutations around the large angle periodic
  trajectory. Our experimental findings are well accounted for by an
  analytical model derived for systems with uniaxial symmetry. They
  also provide new means for controlling highly nonlinear
  magnetization dynamics in nanostructures, which open interesting
  applicative opportunities in the context of magnetic
  nanotechnologies.

\end{abstract}

\maketitle

%%%%%%%%%%%%%%

Spectroscopy based on the resonant interaction of electromagnetic
fields with material media has been of tremendous impact on the
development of physics since the beginning of the 20th century and
remains of crucial importance till nowadays in the study of
nanotechnologies. In this area, a central role is played by magnetic
resonance spectroscopy, which includes various techniques such as
nuclear magnetic resonance (NMR), electron paramagnetic resonance
(EPR), and ferromagnetic resonance (FMR), all based on the excitation
of the Larmor precession of magnetic moments around their equilibrium
position \cite{kittel04}.

FMR differs from NMR and EPR by the fact that in ferromagnetic media,
magnetic moments are coupled by strong exchange interactions which
tend to align them, leading to a large macroscopic spontaneous
magnetization. In these conditions, magneto-dipolar effects become
important and determine large internal fields which enrich both the
ground state, that can be spatially non-uniform, and the dynamics of
magnetic moments. The complex interactions taking place in the media
can be described by an appropriate effective field which sets the time
scale of the magnetization precession, and which itself depends on the
magnetization, making the dynamics, for sufficiently large deviations
from the ground state, highly nonlinear \cite{mayergoyz09}. A special
role in FMR is also played by the spin-waves (SWs), which are the
collective eigenmodes associated to small magnetization oscillations
around the equilibrium configuration \cite{gurevich96}. When pumping
fields excite SWs well above their thermal amplitudes, a rich variety
of phenomena emerges, such as the formation of dynamical solitons
\cite{wigen94, mohseni13}, SW turbulences and chaos \cite{rezende90,
  lvov94, wigen94, petit-watelot12}, and Bose-Einstein condensation of
magnons \cite{demokritov06}, the quanta of SWs.

Recent developments in magnetic nanotechnologies have also
demonstrated that FMR and SW dynamics can be excited either by
microwave magnetic fields or by spin transfer torques, with the
promise of innovative magnonic and spintronic devices for information
and communication technologies \cite{chumak15}.  In this area, spin
torque nano-oscillators (STNOs)
\cite{kiselev03,houssameddine07,chen09,hamadeh12,collet16}, which
exhibit strong nonlinear properties \cite{slavin09}, have even been
successfully implemented to perform neuromorphic tasks
\cite{torrejon17,romera18}.

The complexity of magnetization dynamics when strongly nonlinear
regimes set in is usually detrimental to the reliable control of
nanomagnetic devices, such as oscillators, memories, and logic
gates. In this respect, it is important to establish how far from
equilibrium magnetic nanostructures can be driven before the coherent
magnetization dynamics becomes highly perturbed by the onset of SW
instabilities \cite{suhl57}. In this article, we provide a crucial
advancement in this problem. We demonstrate that FMR in a sufficiently
confined nanostructure can exhibit unprecedented large angle
magnetization precessions which are spatially quasi-uniform. The
experimental evidence of the coherence of large precessions is brought
by a new spectroscopic technique based on the application of a second
probe excitation field, with frequency close to the one of the main
time-harmonic field. This second excitation is used to drive small
eigen oscillations of magnetization around the FMR large angle
periodic oscillations, corresponding to coherent nutations of the
magnetization. These nutation modes are substantially different from
the usual SW modes around the ground state because they correspond to
eigenmodes around a far-from-equilibrium state and their existence is
connected to the one of a large coherent precession. Moreover, we show
that the resonant excitation of these nutations can be used to control
the nonlinear magnetization dynamics by affecting the switching fields
associated with the bistability of the large angle FMR response
pictured in Fig.1a, which occurs beyond the foldover instability
predicted by Anderson and Suhl \cite{anderson55}.

The preservation of coherent magnetization dynamics, which we report
on a thin disc with sub-micrometric diameter, is mainly due to the
geometric confinement. It significantly reduces the density of normal
modes and suppresses the nonlinear SW interactions present in bulk
ferromagnets \cite{kobljanskyj12,melkov13}. In addition, in our
experiments the ground state is with the magnetization perpendicular
to the plane. In this case, the uniform mode in thin films lies at the
bottom of the SW manifold, so that it has no degenerate mode to couple
to \cite{mcmichael90,gulyaev00} and the angle of the purely circular
precession driven by FMR can reach large values
\cite{gnatzig87}. These combined circumstances allow for the
excitation of large-amplitude quasi-uniform precession of
magnetization without simultaneous excitation of other SWs.

The experimental results of the present work also address the
important issue of comparison between theory and experiments. In fact,
by using the framework of dynamical systems \cite{mayergoyz09}, exact
analytical solutions of the Landau-Lifshitz-Gilbert (LLG) equation in
the presence of an arbitrarily large time-harmonic excitation have
been found in high symmetry cases \cite{bertotti01} and their
stability analyzed \cite{bertotti01a}, but have not been verified
experimentally yet. For instance, the hysteretic FMR illustrated in
Fig.1a has already been observed in measurements \cite{fetisov99,
  gui09a}, but with much weaker bistable response characteristics than
expected from theory.

%%%%%%%%%%%%%%

\section{Results}

%%%%%%%%%%%%%%

In the following, we investigate the FMR of an individual nanodisc of
yttrium iron garnet (YIG) in the perpendicular configuration. The
choice of YIG is natural as it is the magnetic material with the
smallest SW damping, making it attractive to study weakly dissipative
magnetization dynamics in the linear and nonlinear regimes
\cite{serga10}. The nanodisc has a diameter of 700~nm and is patterned
from a 20~nm thick YIG film of magnetization $\mu_0 M_s = 0.21$~T
\cite{kelly13}, $\mu_0$ being the vacuum permeability. It is saturated
out-of-plane by a magnetic field $\bm H_0$ applied along its normal,
$z$. A broadband antenna supplies a spatially uniform, linearly
polarized microwave field of pulsation $\omega_1$ oriented in the
plane of the nanodisc. It can be decomposed into the left and right
circularly polarized components, only the latter being efficiently
coupled to the Larmor precession of the magnetization. In the
following, $h_1$ will refer to the \emph{circular} amplitude of the
excitation field produced by the output power $P_1$ from the
synthesizer. It drives the YIG nanodisc into FMR, thereby opening a
precession angle $\theta$ of the magnetization $\bm M$ around $\bm
H_0$ and decreasing its longitudinal component $M_z = M_s
\cos\theta$. This dynamics is characterized by magnetic resonance
force microscopy (MRFM), which sensitively probes the variation
$\Delta M_z = M_s - M_z$ through the dipolar force between the YIG
nanodisc and a magnetic nanosphere attached at the end of a soft
cantilever \cite{klein08}, as sketched in Fig.1b. Further details
about the sample, the MRFM set-up and the microwave calibration can be
found in the Methods.

%%%%%%%%%%%%%%

\begin{figure}
  \includegraphics[width=12.5cm]{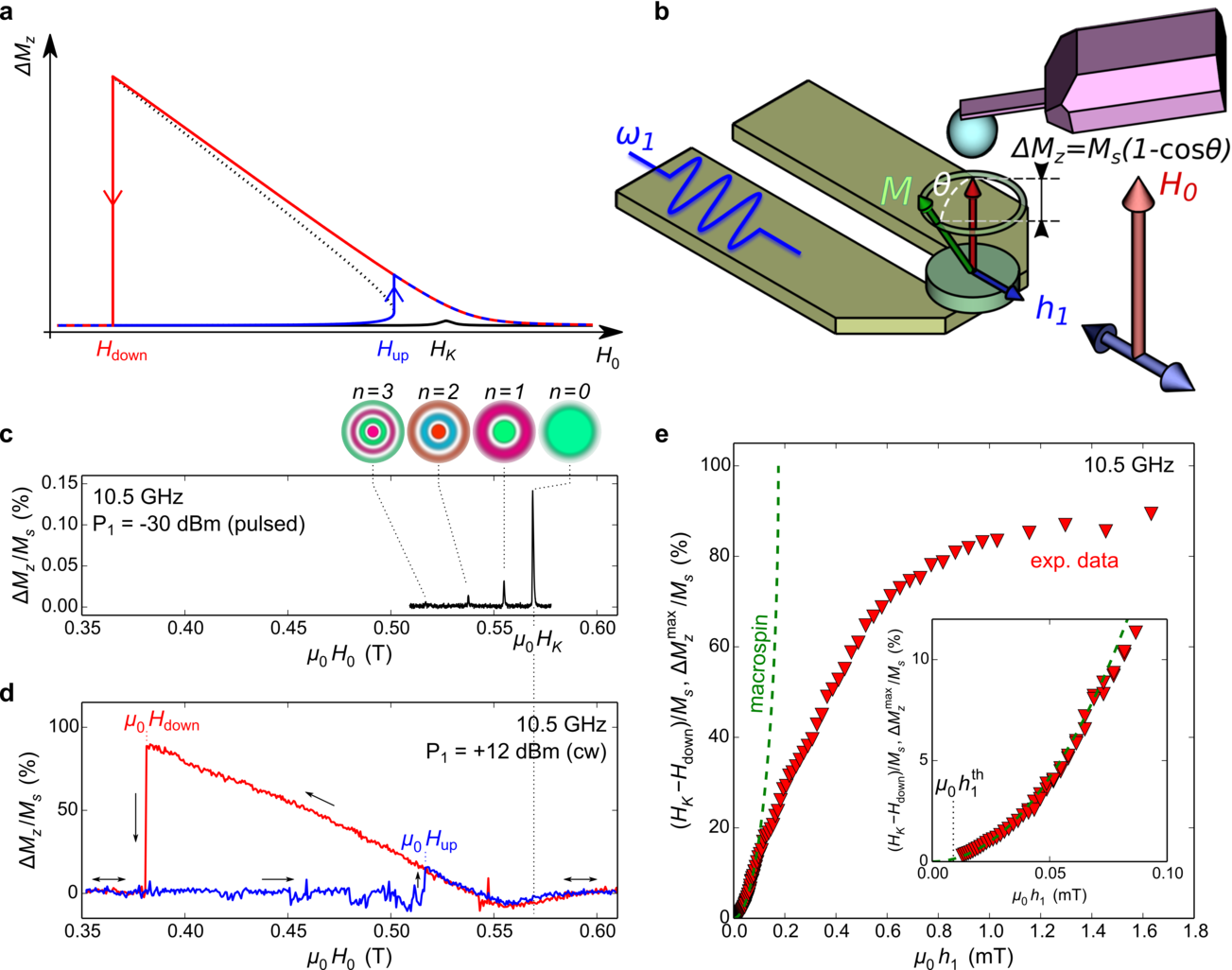}
  \caption{\textbf{Ultrastrong foldover of FMR.} (a) Illustration of
    hysteretic foldover in the nonlinear regime of FMR, where jumps
    between the two stable branches of the dynamics occur at
    $H_\text{down}$ and $H_\text{up}$. The dotted line is the unstable
    branch. The black Lorentzian curve centered at $H_K$ corresponds
    to linear FMR. (b) Schematics of the experiment. A microwave field
    $\bm h_1$ of pulsation $\omega_1$ drives the magnetization $\bm M$
    of a YIG nanodisc into FMR, opening a precession angle $\theta$
    around the perpendicularly applied field $\bm H_0$. The associated
    variation in the longitudinal component of the magnetization,
    $\Delta M_z$, is mechanically detected by the cantilever of a
    magnetic resonance force microscope. (c) Spin-wave spectroscopy
    performed at 10.5~GHz in the linear regime. The profiles of the
    quantized radial SW modes, calculated using a micromagnetic code,
    are shown above (different colors display regions precessing in
    opposite phase). (d) FMR spectrum in the deeply nonlinear regime
    exhibiting ultrastrong hysteretic foldover and nearly complete
    suppression of $M_z$. (e) Evolution of the maximal normalized
    foldover shift as a function of the pumping field $h_1$. The
    dashed line shows the behavior expected for a macrospin and the
    inset a zoomed view of the low amplitude regime.}
  \label{FIG1}
\end{figure}

%%%%%%%%%%%%%%

\subsection{Linear spin-wave spectroscopy}

In Fig.1c, the SW spectroscopy of the YIG nanodisc is performed at
$\omega_1/(2\pi) = 10.5$~GHz and low power $P_1 = -30$~dBm, which is
pulse modulated at the frequency of the MRFM cantilever to improve the
signal to noise ratio. Quantized radial SW modes are excited by the
uniform pumping field \cite{naletov11}. Their spatial profiles indexed
by the radial number are shown above the spectrum. The fundamental
Kittel mode is the one excited at the largest field, $\mu_0H_K =
0.569$~T, and corresponds to a uniform phase of the transverse
magnetization in the disc. Due to the geometric confinement, it is
well separated from other SW modes at lower field \cite{hahn14}. Its
full width at half maximum, $\mu_0 \Delta H = 0.35$~mT, is determined
at even lower power ($P_1 = -38$~dBm) to avoid distortions of the
resonance line due to the onset of foldover, which occurs when the
change in effective field becomes comparable to the FMR linewidth, at
$P_1^\text{th} = -33$~dBm or $\mu_0 h_1^\text{th} = 0.009$~mT (see
Methods). It corresponds to a Gilbert damping parameter $\alpha =
\gamma \mu_0 \Delta H / (2 \omega_1) = 4.7 \cdot 10^{-4}$, where
$\gamma$ is the gyromagnetic ratio, in agreement with the value
determined from broadband measurements (see Supplementary Fig.S1).

%%%%%%%%%%%%%%

\subsection{Deeply nonlinear FMR}

The FMR spectrum of the YIG nanodisc radically changes at much
stronger pumping fields. Fig.1d shows the measurement with a
continuous wave (cw) excitation at $P_1 = +12$~dBm, \textit{i.e.},
more than four orders of magnitude larger than the threshold of
foldover instability.  The cw excitation allows to reveal the bistable
character of the nonlinear magnetization dynamics. By sweeping down
the applied field (red curve) through the resonance of the Kittel
mode, the precession angle substantially increases, which decreases
the static demagnetizing field $\mu_0 M_z$ and shifts the FMR
condition $\omega_1 = \gamma \mu_0 (H_0 - M_z)$ to lower magnetic
field by the same amount. This foldover shift to lower field continues
until the pumping field cannot sustain anymore the large-amplitude
magnetization dynamics, causing the sharp downward jump to the lower
stable branch observed at $\mu_0 H_\text{down}=0.381$~T. By sweeping
up the applied field (blue curve), an upward jump to the higher stable
branch is observed at $\mu_0 H_\text{up}=0.516$~T. The extremely
hysteretic foldover witnessed in this experiment is
remarkable. Moreover, the maximal foldover shift
$\mu_0(H_K-H_\text{down}) = 0.188$~T corresponds to a reduction of
nearly 90\% of $\mu_0M_z$ induced by the microwave pumping, which
translates into a mean precession angle of 84$^\circ$ in the
nanodisc. The evolution of the maximal normalized foldover shift as a
function of the pumping field $h_1$ is plotted in Fig.1e together with
$\Delta M_{z}^\text{max} / M_s = 1 - \sqrt{1 - 4h_1^2/\Delta H^2}$
calculated from the macrospin LLG equation \cite{gurevich96}. The
measured foldover shift starts to deviate from the macrospin model
beyond $\mu_0 h_1 \simeq 0.1$~mT, which is an order of magnitude
larger than the threshold for foldover, when the angle of the uniform
precession increases above 30$^\circ$, corresponding to $\Delta
M_z/M_s \simeq 15$\%. This is the signature of the onset of SW
instabilities \cite{suhl57}, which is here significantly postponed
compared to what is observed in larger YIG samples, where the Suhl
threshold is reached even before the onset of foldover, for a uniform
precession angle of only a couple of degrees
\cite{loubens05,naletov03}. In this respect, the experimental results
presented in Fig.1 demonstrate that the discretization of the
excitation spectrum in nanostructures efficiently inhibits nonlinear
interactions between SW modes, which drastically modifies the high
amplitude magnetization dynamics \cite{kobljanskyj12,melkov13}.

%%%%%%%%%%%%%%

\begin{figure}
  \includegraphics[width=12.5cm]{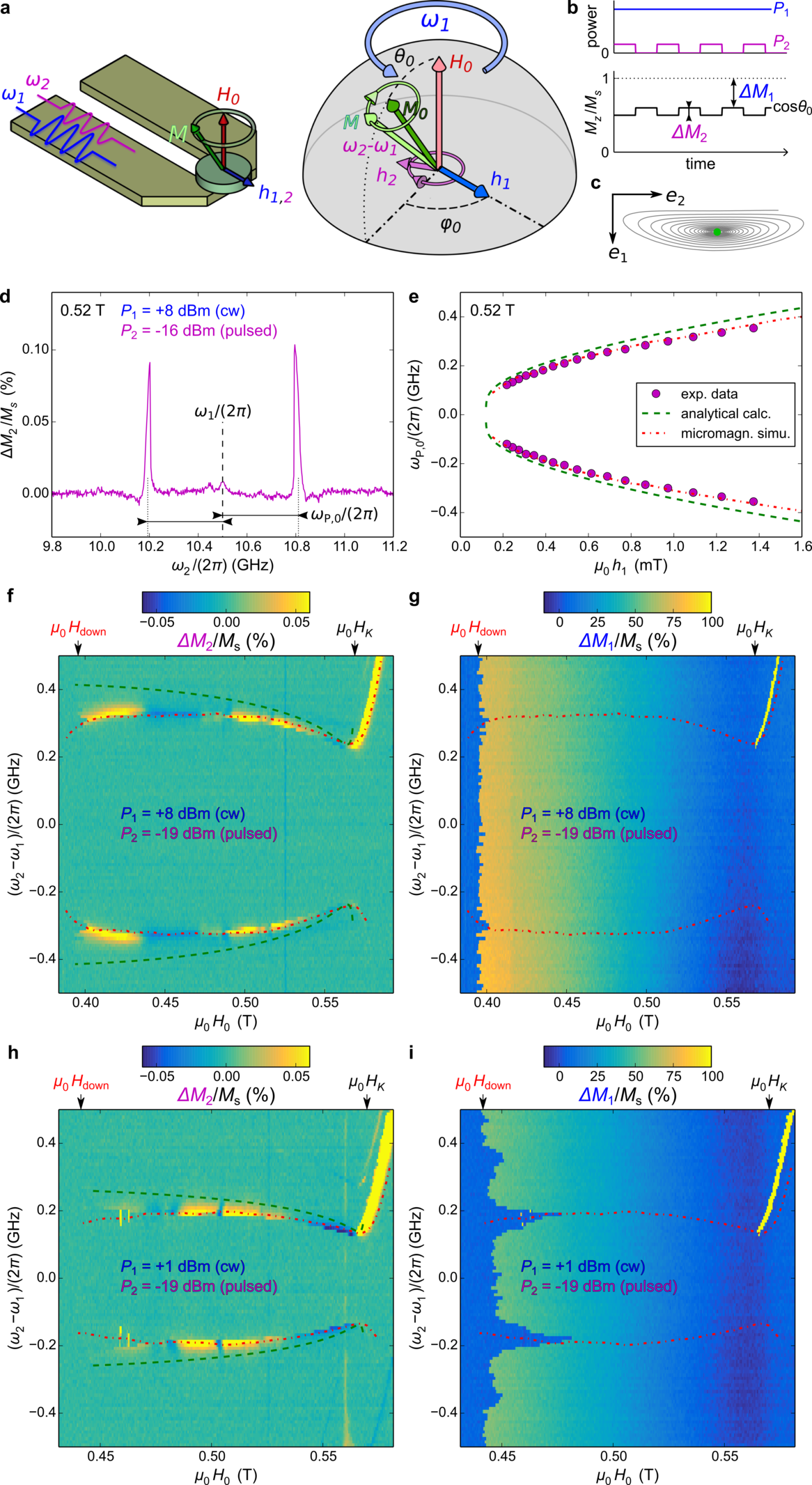}
  \caption{\textbf{Nutation of magnetization.} (a,b) Principle of the
    experiment. A low power pulse modulated microwave field $\bm h_2$
    of pulsation $\omega_2$ is added to the main pumping cw field $\bm
    h_1$. This enables the spectroscopy in the frame rotating with
    $\bm h_1$ at $\omega_1$, where the magnetization $\bm M_0$ of the
    \textbf{P}-mode is fixed at an angle $\theta_0$ and a phase lag
    $\varphi_0$. (c) Relaxation trajectory of the magnetization
    towards the \textbf{P}-mode in the plane $(\bm e_1,\bm e_2)$
    orthogonal to $\bm M_0$ defined in the Methods, calculated using
    micromagnetic simulations. (d) Spectroscopy performed at
    $\omega_1/(2\pi)=10.5$~GHz and $\mu_0 H_0=0.52$~T as a function of
    $\omega_2$. The two resonance peaks symmetric with respect to
    $\omega_1$ correspond to a motion of nutation in the laboratory
    frame. (e) Evolution of the nutation frequency as a function of
    the main pumping field $h_1$ at fixed $\mu_0 H_0=0.52$~T. (f)
    Nutation spectroscopy map at fixed $h_1$ of the small amplitude
    dynamics $\Delta M_2$ excited by $\omega_2$ as a function of the
    down swept field $H_0$. (g) Simultaneous measurement of $\Delta
    M_1$ induced by the main pumping at $\omega_1$. (h,i) Same as
    (f,g) for a smaller main pumping power. In panels (e)--(i), the
    green dashed lines show the analytical predictions from
    Eq.\ref{eq-nutfreq} and red dotted lines the results from
    micromagnetic simulations.}
  \label{FIG2}
\end{figure}

%%%%%%%%%%%%%%

\subsection{Nutation spectroscopy in the rotating frame}

We now aim at probing the stability of the large-amplitude
magnetization dynamics demonstrated above, which is periodic at
$\omega_1$ in the laboratory frame, hence referred to as
\textbf{P}-mode \cite{bertotti01}. In the frame rotating with $\bm
h_1$ at $\omega_1$ around the $z$-axis, the magnetization $\bm M_0$ of
a \textbf{P}-mode is fixed at a polar angle $\theta_0$ and a phase lag
$\varphi_0$ (see Fig.2a and supplementary information). For this, we
conduct two-tone measurements, where in addition to the strong cw
excitation $\bm h_1$ at $\omega_1$ a second weak microwave field $\bm
h_2$, pulse modulated at the cantilever frequency, is applied at
$\omega_2$, as shown in Fig.2b. MRFM is used to simultaneously detect
$\Delta M_1$ induced by the main cw pumping at $\omega_1$ (by
monitoring the cantilever frequency, as in Fig.1d) and the additional
change in longitudinal magnetization, $\Delta M_2$, induced by the
second excitation at $\omega_2$ (by monitoring the amplitude of the
cantilever vibrations, as in Fig.1c). The former informs us about the
time-harmonic steady state regime driven by $\bm h_1$, whereas the
latter allows us to spectroscopically probe the eigen excitations on
top of this \textbf{P}-mode. The $\Delta M_2$ spectrum measured at
constant bias field $\mu_0H_0=0.52$~T by sweeping $\omega_2/(2\pi)$ at
low power $P_2=-16$~dBm in the vicinity of the frequency
$\omega_1/(2\pi)=10.5$~GHz of the main pumping ($P_1=+8$~dBm) is shown
in Fig.2d. It displays two narrow resonance peaks centered at 10.2~GHz
and 10.8~GHz, \textit{i.e.}, symmetrically with respect to
$\omega_1/(2\pi)$. This means that in the frame rotating with $\bm
h_1$ at $\omega_1$, the magnetization is precessing at
$(\omega_2-\omega_1)/(2\pi) = \pm \omega_{\text{P},0}/(2\pi) = \pm
0.3$~GHz around its equilibrium position $\bm M_0$ (cf. Fig.2a). In
other words, it is submitted to a slow nutation motion in the
laboratory frame. The dependence of the nutation frequency
$\omega_{\text{P},0}$ on the main pumping field $h_1$ at fixed
$\mu_0H_0=0.52$~T is presented in Fig.2e, whereas its evolution
measured as a function of the down swept field $H_0$ at fixed
$P_1=+8$~dBm and $P_1=+1$~dBm is shown in the 2D spectroscopy maps of
Fig.2f and 2h, respectively.

Following the theoretical approach of ref.\cite{bertotti01a}, it is
possible to calculate analytically the frequency $\omega_{\text{P},0}$
of spatially uniform nutation around a given \textbf{P}-mode based on
the macrospin LLG equation. Technical details are given in the Methods
(the full derivation is presented in the supplementary
information). In the limit of small damping, $\alpha \ll 1$, it can be
expressed as a function of $h_1$ and the angles $\theta_0$ and
$\varphi_0$ of the \textbf{P}-mode as follows:
\begin{equation}
  \frac{\omega_{\text{P},0}^2}{\gamma^2} = \frac{\mu_0h_1\cos\varphi_0}{\sin\theta_0}\left( \frac{\mu_0h_1\cos\varphi_0}{\sin\theta_0}+ \mu_0M_s\sin^2\theta_0 \right) \, .
\label{eq-nutfreq}
\end{equation}
This analytical expression is plotted as green dashed lines in
Figs.2e, 2f and 2h, using the amplitude $h_1$ of the main driving
field in the experiments, the angle of precession at the bias field
$H_0$ determined from the normalized foldover shift, $\cos\theta_0
=(H_K-H_0)/M_s$, and the phase lag which satisfies $\gamma
\mu_0h_1\sin\varphi_0 = \alpha \omega_1 \sin\theta_0$ in the macrospin
model. It reproduces rather well the experimental data, except in
regions where the level of excitation is very large, due to the
deviation from the macrospin behavior already observed in Fig.1e. In
addition, we have conducted full micromagnetic simulations in the time
domain (see Methods), which allow us to extract the nutation frequency
from the relaxation of the magnetization towards the steady state
regime driven by $\bm h_1$, shown in Fig.2c. The obtained results,
plotted as red dotted lines in Figs.2e, 2f and 2h, quantitatively
agrees with the data on the full range of parameters investigated.

The dependence of the nutation frequency on $h_1$ observed in Fig.2e
can be explained as follows. There is a minimum amplitude of $\mu_0h_1
\simeq 0.15$~mT to drive the \textbf{P}-mode at $\mu_0H_0=0.52$~T,
which corresponds to a normalized foldover shift $(H_K-H_0)/M_s =
23$\%, \textit{i.e.}, an angle of precession $\theta_0 \simeq
40^\circ$. Above this amplitude, the nutation frequency is defined and
increases with $h_1$ as predicted by Eq.\ref{eq-nutfreq}, which shows
that there are two torques driving the nutation dynamics of the
magnetization. The first one is linear, and directly provided by $\bm
h_1$, which sets the Rabi frequency in a magnetic resonance experiment
\cite{capua17}. The second one is a demagnetizing torque specific to
nonlinear FMR, which stiffens the nutation resonance.

%%%%%%%%%%%%%%

\begin{figure*}
  \includegraphics[width=17.5cm]{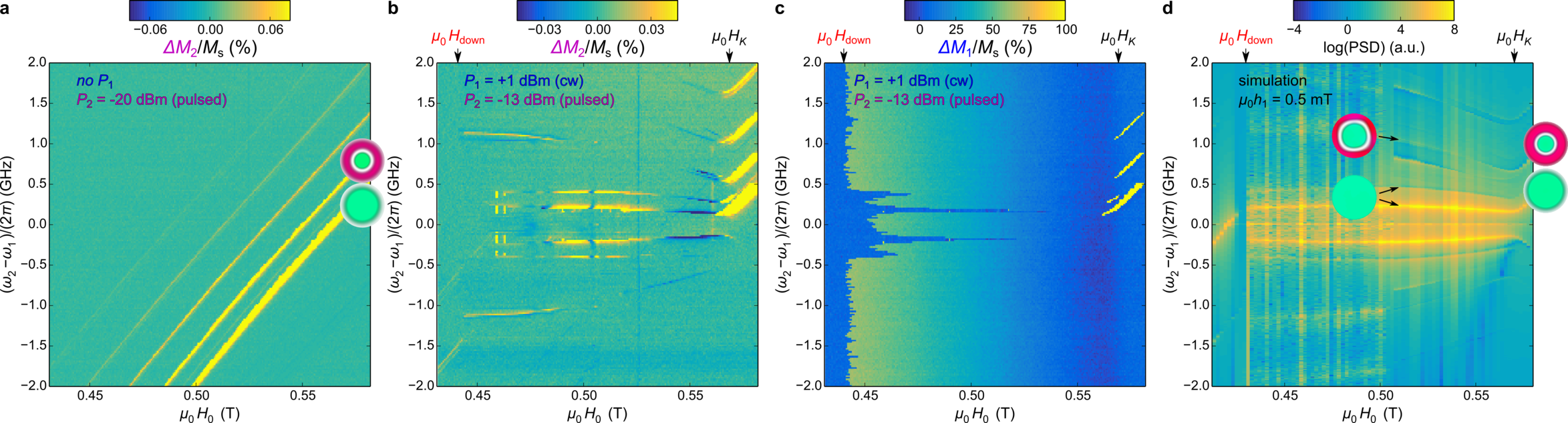}
  \caption{\textbf{Spin-wave nutation modes.} (a) Spectroscopy map of
    the SW modes excited by the low power excitation at $\omega_2$
    \emph{in the absence} of the main pumping at
    $\omega_1/(2\pi)=10.5$~GHz. They correspond to the same radial
    modes as probed in the linear regime in Fig.1c, whose profiles are
    recalled as insets. (b) $\Delta M_2$ spectroscopy map of the SW
    nutation modes excited by the low power excitation at $\omega_2$
    \emph{in the presence} of the main excitation at $\omega_1$
    ($P_1=+1$~dBm). (c) Simultaneous measurement of $\Delta M_1$
    induced by the main pumping at $\omega_1$. (d) Micromagnetic
    simulations of the experimental data shown in (b). The SW mode
    profiles shown as insets are extracted at some specific
    $\omega_2-\omega_1$ and $H_0$. In all the panels, the
    perpendicular field $H_0$ is swept down.}
  \label{FIG3}
\end{figure*}

%%%%%%%%%%%%%%

We now interpret the nutation spectroscopy maps of Figs.2f and 2h,
where $\Delta M_2$ is measured at fixed $h_1$ by sweeping $H_0$ and
$\omega_2$. When $H_0 > H_K$, the magnetization dynamics is driven off
resonantly by $\bm h_1$ and has small amplitude. Hence, the weak
microwave field $\bm h_2$ simply excites the linear Kittel resonance
on top of it, which explains the bright linear dispersion $\omega_2 =
\gamma \mu_0 (H_0 - M_s)$ observed in this region. The situation is
quite different when $H_0 < H_K$. In this case, $\bm h_1$ drives the
strong foldover regime demonstrated in Fig.1 upon sweeping down $H_0$,
and $\bm h_2$ excites the magnetization dynamics on top of the
corresponding \textbf{P}-mode. The evolution of the two resonance
branches symmetrically distributed around the main pumping frequency
$\omega_1$ as a function of $H_0$ is reproduced qualitatively by
Eq.\ref{eq-nutfreq}, and quantitatively by micromagnetic
simulations. The fact that the upper branch is continuous with the
linear Kittel resonance branch observed above $H_K$ indicates that the
perturbation of the \textbf{P}-mode driven by $\bm h_2$ has a uniform
phase, \textit{i.e.}, corresponds to a uniform nutation of the
magnetization.

The experimental results presented in Fig.2 also demonstrate that the
weak resonant excitation of the nutation mode can destabilize the
strong foldover dynamics. Figs.2g and 2i display the evolution of
$\Delta M_1$ induced by the cw pumping $\bm h_1$ at $\omega_1$, while
exciting the nutation dynamics with $\bm h_2$ as a function of $H_0$
and $\omega_2$, whose $\Delta M_2$ spectroscopy is presented in
Figs.2f and 2h, respectively. In these 2D maps, the foldover breakdown
occurring at $H_\text{down}$ is easily identified thanks to the
associated sharp change of $\Delta M_1$, and is anticipated as the
nutation mode is excited. This is particularly clear in Fig.2i, where
the maximal foldover shift $\mu_0(H_K-H_\text{down})$ is reduced by
almost 0.05~T when $|\omega_2-\omega_1|/(2\pi) \simeq 0.2$~GHz, which
corresponds to the nutation resonance. Moreover, these data suggest
that higher order nutation modes can be excited by $\bm h_2$, since an
anticipated foldover breakdown is also observed at
$|\omega_2-\omega_1|/(2\pi) \simeq 0.35$~GHz.

In order to investigate these other nutation modes, we perform the
same measurements as in Figs.2f--i, but for larger detunings
$\omega_2-\omega_1$ ($P_2$ is also increased from $-19$ to
$-13$~dBm). The results obtained at $P_1=+1$~dBm are reported in
Figs.3b--c (those obtained at $P_1=+8$~dBm are presented in
Supplementary Fig.S2). The spectroscopy map of the SW modes excited by
$\bm h_2$ in the absence of $\bm h_1$ is shown in Fig.3a. It displays
the linear dispersion relation of the radial SW modes excited by the
uniform field $\bm h_2$, already discussed in Fig.1c. Due to the
strong foldover regime driven by $\bm h_1$ at $H_0<H_K$ in Fig.3b,
each of these radial SW branches transforms into a pair of branches
symmetric around $\omega_1$. Additionally, there is a pair of branches
which appears at twice the main nutation frequency, which is due to
the ellipticity of the trajectory, apparent on Fig.2c. The macrospin
approach used to derive Eq.\ref{eq-nutfreq} cannot be used to account
for these higher order nutation modes, although plane wave
perturbations to the \textbf{P}-mode can also be analytically
calculated \cite{bertotti01a}. We therefore use micromagnetic
simulations to calculate the SW nutation spectra shown in Fig.3d,
which are in good agreement with the experiments. The extracted SW
nutation modes profiles (see Methods), shown as insets in Fig.3d,
indicate that there is some continuity between the radial SW modes
excited in the linear regime on top of the equilibrium magnetization
and the nutation modes excited on top of the \textbf{P}-mode driven in
the nonlinear regime by $\bm h_1$. Finally, Fig.3c confirms that the
excitation of the nutation resonances can destabilize the strong
foldover dynamics.

%%%%%%%%%%%%%%

\section{Discussion}

As in the case of a spinning top, the nutation of magnetization
demonstrated above is made possible by the specific properties of the
dynamics on the unit sphere \cite{mayergoyz09}. Namely, it is
topologically allowed for the magnetization to oscillate around its
fixed point $\bm M_0$ $(\theta_0,\varphi_0)$ in the rotating frame,
which is set by the drive $\bm h_1$. The nutation frequency results
from the balance of torques acting on the angular momentum, and is
given by Eq.\ref{eq-nutfreq} in the case of a macrospin governed by
the LLG equation. The accuracy of the latter to account for the
experimental data means that the coherent precession of the
magnetization vector is dominating the deeply nonlinear driven
dynamics, despite the signatures of SW instabilities observed at very
large pumping power. Their main effect is to slightly reduce the
nutation frequency, which is well captured by full micromagnetic
simulations. This can be ascribed to the shift of the phase between
the pumping field and the average magnetization \cite{bauer15}
observed in our simulations, a key effect to explain the above
threshold dynamics \cite{lvov94,gerrits07}. The nutation spectroscopy
of magnetization thus allows a more detailed investigation of the
highly nonlinear regime, where auto-oscillation instabilities
\cite{rezende90,mcmichael90,luehrmann91,watanabe17} and instability
patterns \cite{bonin12} have been evidenced. Moreover, its
quantitative understanding, made possible thanks to the low density of
SW modes in nanomagnets, should enable to test further the LLG
equation governing the motion of magnetization against experimental
measurements.

Our results also highlight that the dynamical states driven by a high
power microwave signal can be controlled using a second signal with
much lower power by the resonant excitation of the nutation
modes. This could be applied in devices taking advantages of the
bistable magnetization dynamics for microwave signal processing
\cite{fetisov99}, in analogy to microwave assisted magnetization
switching \cite{thirion03, pigeau11, suto18}. Furthermore, the
frequency selectivity and energy efficiency of nutation excitations
provide new potentials for the scheme of neuromorphic
computing. Cognitive tasks have already been implemented using the
nonlinear dynamics of nanomagnets, from the transient regime of a
single STNO \cite{torrejon17} to the collective behavior of mutually
coupled STNOs controlled by external microwave signals
\cite{romera18,li17}. An appropriate use of the nutation dynamics of
magnetization would allow to gain additional control on nonlinear
dynamics, which is highly desired in this field.

%%%%%%%%%%%%%%

\section{Methods}

\textbf{Sample preparation.} A 20~nm thick Y$_3$Fe$_5$O$_{12}$ (YIG)
film was grown by pulsed laser deposition on a (111)
Gd$_3$Ga$_5$O$_{12}$ (GGG) substrate, as described in
ref.\cite{kelly13}. It was used to pattern the studied YIG nanodisc by
electron lithography and dry etching. After the insertion of a 50~nm
thick SiO$_2$ insulating layer, a 150~nm thick and 5~$\mu$m wide gold
antenna was defined on top of the nanodisc to provide the microwave
excitation \cite{hahn14}.

\textbf{MRFM set-up.} The magnetic resonance force microscope is
located between the poles of an electromagnet and operated under
vacuum (10$^{-6}$ mbar) at a stabilized temperature of 288~K. The
cantilever is an Olympus Biolever (spring constant
5~mN$\cdot$m$^{-1}$) with a 700-nm-diameter sphere of an amorphous
FeSi alloy (magnetic moment 0.28~pA$\cdot$m$^2$) glued to its apex. In
this study, MRFM spectroscopy is achieved by placing the center of
this magnetic nanosphere at a distance of 1.5--1.8~$\mu$m above the
center of the YIG nanodisc. The strayfield of the MRFM probe
(10--16~mT) is subtracted from the corresponding spectra. The
displacement of the cantilever is monitored using optical
techniques. Its mechanical frequency ($f_c \approx 12.3$~kHz) is
tracked using a phase-locked loop and its vibration amplitude
stabilized to 4~nm using a piezoelectric bimorph. When the cw
microwave pumping excites the magnetization dynamics in the sample,
its longitudinal component is reduced, so the static dipolar force
with the magnetic probe diminishes. The associated variation of the
cantilever frequency provides a quantitative magnetometry of the
sample \cite{lavenant14}. In order to improve the signal to noise
ratio, the microwave excitation is pulsed on and off at $f_c$. In that
case, the cantilever vibrations induced by the magnetization dynamics
excited in the sample are enhanced by the quality factor $Q \approx
2000$ of the mechanical detection \cite{klein08}.

\textbf{Microwave field calibration.} We use the onset of foldover as
a mean to calibrate the amplitude of the excitation field produced by
the microwave antenna at the sample location \cite{naletov11}. At the
threshold of foldover instability $h_1^\text{th}=0.62 \Delta
H^{3/2}/M_s^{1/2}$ \cite{anderson55}, and the slope of the FMR curve
becomes infinite on the low field side of the resonance, which is
experimentally observed at 10.5~GHz for an output power from the
synthesizer $P_1^\text{th} = -33$~dBm. Using the FMR line width
measured in the linear regime, one gets $\mu_0 h_1^\text{th} =
0.009$~mT, \textit{i.e.}, a calibration factor $a =
0.4$~mT$/\sqrt{\text{mW}}$ between microwave field and power. To get a
better precision, we also fit the dependences on power of the critical
fields $H_\text{down}$ and $H_\text{up}$ determined experimentally
beyond the foldover onset \cite{gui09a}, which yields $a=0.41 \pm
0.03$~mT$/\sqrt{\text{mW}}$.

\textbf{Analytical calculations.} The nonlinear FMR excited in a
uniaxial system by the superposition of two time-harmonic external
fields, $\bm h_\text{ac}(t) = \bm h_1(t) + \bm h_2(t)$ with $| \bm
h_2(t)| \ll | \bm h_1(t) | $, is calculated based on the macrospin LLG
equation. The main stages of the analytical derivation presented in
the supplementary information are the following. The LLG equation is
first written in the frame of reference rotating around the $z$-axis
at the angular frequency $\omega_1$ of the dominant time-harmonic
component: $ \dot{\bm m} -\alpha \bm m \times\dot{\bm m} = -\bm m
\times \left( \bm h_\text{eff} -\omega_1 \bm e_z\right) +\alpha
\omega_1 \bm m \times(\bm e_z \times \bm m) $ where $ \bm h_\text{eff}
= \kappa_\text{eff} m_z \bm e_z + h_0 \bm e_z +\bm h_\text{ac}(t)$,
$\dot{\bm m}$ is the time derivative of the normalized magnetization
vector taken in the rotating frame, $\bm h_0$ the normalized bias
field, $\bm e_z$ the unit vector along it, and $\kappa_\text{eff}$ the
effective anisotropy constant. It is then written in spherical
coordinates and considered in the case where only the right circularly
polarized component of $\bm h_1$ is applied. This allows to find its
equilibrium points $\bm m_0$ (\textbf{P}-modes), or equivalently
$(\theta_0,\varphi_0)$, and to analyze the foldover of FMR
\cite{bertotti01}. The standard analysis of the stability of these
\textbf{P}-modes allows to calculate the nutation frequency given in
Eq.\ref{eq-nutfreq} \cite{bertotti01a}.  Shortly, $\bm m$ is expanded
around $\bm m_0$ to linearize the LLG equation, and the complex
amplitudes of magnetization perturbations are calculated in the
rotating frame by projecting them in the plane $(\bm e_1, \bm e_2)$
orthogonal to $\bm m_0$, where $\sin\theta_0 \bm e_1 = (\bm e_z \times
\bm m_0)\times \bm m_0 $ and $\sin\theta_0 \bm e_2 = (\bm e_z \times
\bm m_0)$. Finally, the \textbf{P}-mode linear response to the small
additional microwave field $\bm h_2$ is studied.

\textbf{Micromagnetic simulations.} The magnetization dynamics in the
YIG nanodisc is calculated using the python module MicroMagnum, a
micromagnetic finite difference simulator which can be runned on GPU
\cite{magnum}. The nominal geometry of the nanodisc (diameter 700~nm,
thickness 20~nm) is discretized using a $128\times128\times1$
rectangular mesh. The following magnetic parameters are used: $M_s =
1.67 \cdot 10^5 $~A$\cdot$m$^{-1}$, $A_\text{ex} = 4.3 \cdot
10^{-12}$~J$\cdot$m$^{-1}$ (exchange length $\simeq 16$~nm),
$\gamma/(2\pi) = 28.5$~GHz$\cdot$T$^{-1}$$\cdot$s$^{-1}$, and $\alpha
= 5 \cdot 10^{-4}$. The bias magnetic field is applied along the
normal of the disc. The static equilibrium configuration of the
magnetization is calculated at 0.59~T. Then a linearly polarized
excitation field of constant amplitude is applied at 10.5~GHz in the
plane of the disc, and for each value of the bias magnetic field,
which is decreased by steps of 0.02~T, the resulting magnetization
dynamics is calculated over 100~ns with a typical step of 3~ps. This
allows to reproduce the foldover regime demonstrated in the
experiments and to calculate the nutation frequencies. Those are
obtained by fast Fourier transformation of the transient dynamics of
the average magnetization simulated at each bias field. The nutation
mode profiles are obtained by stroboscopically averaging the
magnetization dynamics at the corresponding nutation frequencies in
the rotating frame.

%%%%%%%%%%%%%%

\section{Acknowledgements}

Y.L. thanks the French National Research Agency for funding under
contract No. ANR-14-CE26-0021 (MEMOS), V.V.N. the Russian Competitive
Growth program of KFU, and C.S. the Labex NanoSaclay for support under
the action ``Rayonnement International''. This research was partially
funded by the ANR-18-CE24-0021 (MAESTRO) project.

%%%%%%%%%%%%%%

% \section{Author contributions}

% All authors discussed the results and commented on the manuscript.

%%%%%%%%%%%%%%

% \section{Supplementary information}

% The authors declare no competing financial interests. Correspondence
% and requests for materials should be addressed to G.d.L.

%%%%%%%%%%%%%%

\begin{figure}
  \includegraphics[width=12.5cm]{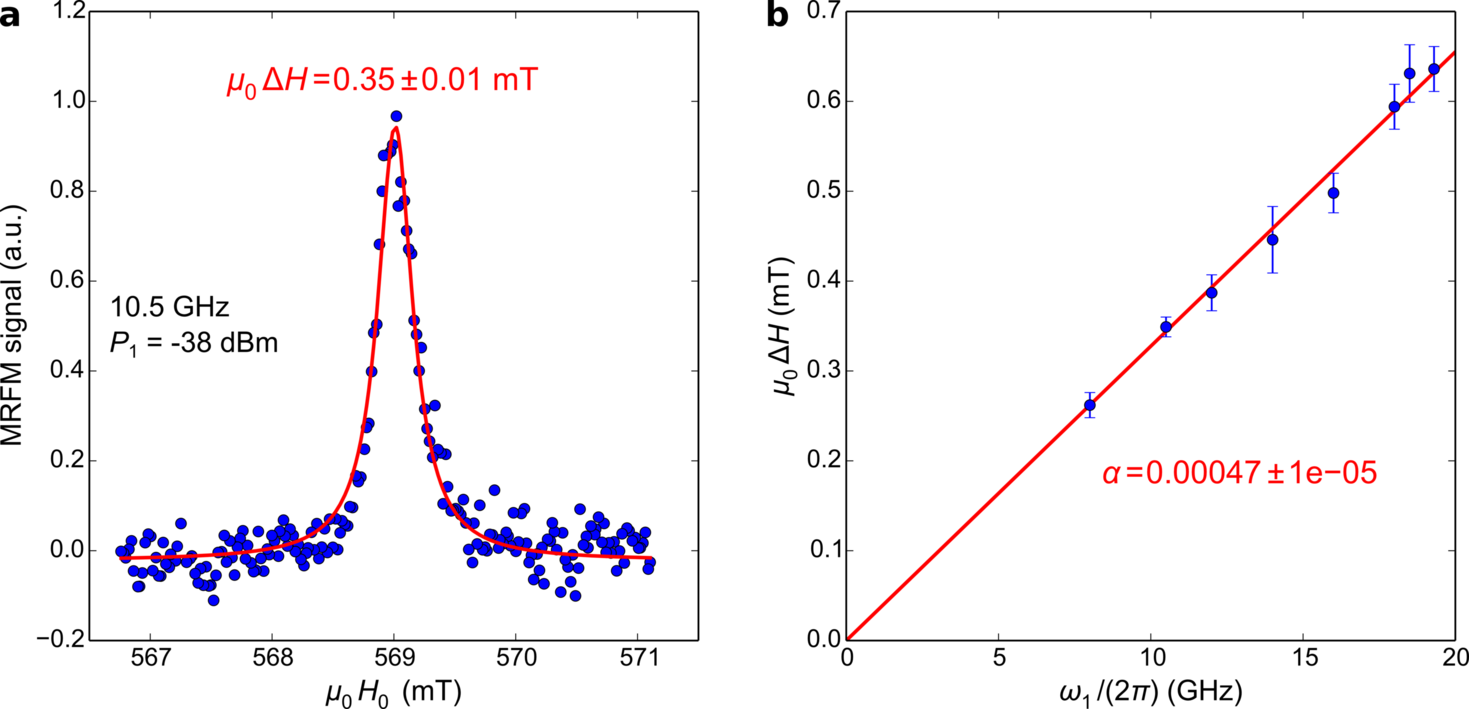}
  \caption{(Supplementary Fig.S1) \textbf{Gilbert damping of the YIG
      nanodisc.} (a) FMR peak measured at $\omega_1/(2\pi)=10.5$~GHz
    and $P_1=-38$~dBm. A Lorentzian fit to the data yields the full
    width at half maximum $\mu_0\Delta H$ together with the associated
    error bar. (b) Dependence on excitation frequency of $\mu_0\Delta
    H$ determined in the linear regime. A linear fit to the data
    yields the Gilbert damping parameter of the YIG nanodisc.}
  \label{figsup}
\end{figure}

\begin{figure*}
  \includegraphics[width=17cm]{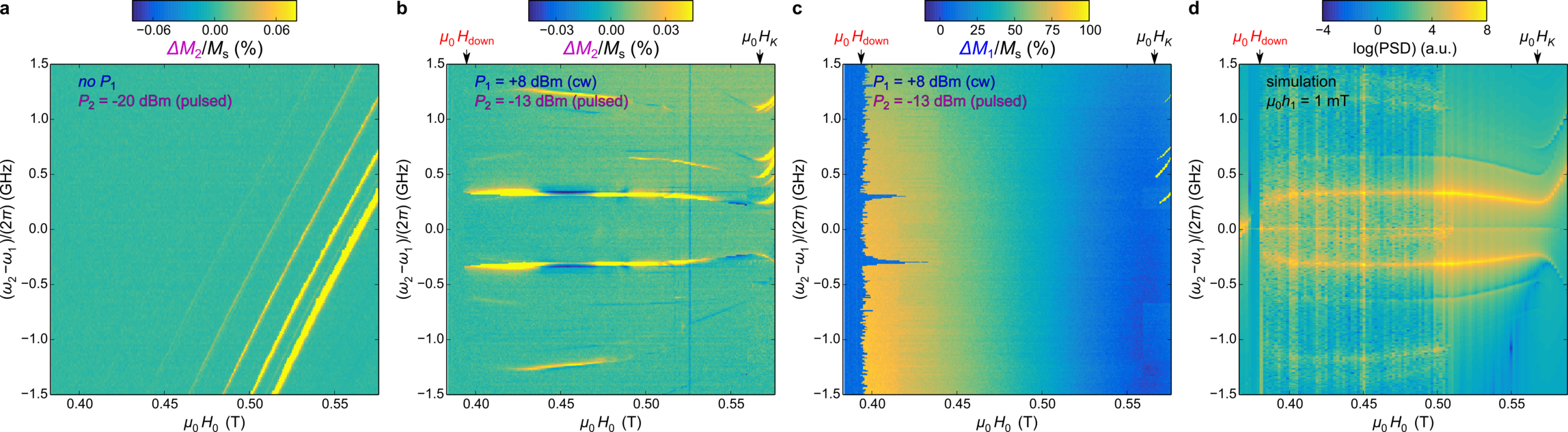}
  \caption{(Supplementary Fig.S2) \textbf{Spin-wave nutation modes at
      larger main pumping power.} (a) Spectroscopy map of the SW modes
    excited by the low power excitation at $\omega_2$ \emph{in the
      absence} of the main pumping at $\omega_1/(2\pi)=10.5$~GHz. (b)
    $\Delta M_2$ spectroscopy map of the SW nutation modes excited by
    the low power excitation at $\omega_2$ \emph{in the presence} of
    the main excitation at $\omega_1$ ($P_1=+8$~dBm). (c) Simultaneous
    measurement of $\Delta M_1$ induced by the main pumping at
    $\omega_1$. (d) Micromagnetic simulations of the experimental data
    shown in (b). In all the panels, the perpendicular field $H_0$ is
    swept down.}
  \label{figsup2}
\end{figure*}

%%%%%%%%%%%%%%

%

\newpage


\begin{thebibliography}{0}%
\makeatletter
\providecommand \@ifxundefined [1]{%
 \@ifx{#1\undefined}
}%
\providecommand \@ifnum [1]{%
 \ifnum #1\expandafter \@firstoftwo
 \else \expandafter \@secondoftwo
 \fi
}%
\providecommand \@ifx [1]{%
 \ifx #1\expandafter \@firstoftwo
 \else \expandafter \@secondoftwo
 \fi
}%
\providecommand \natexlab [1]{#1}%
\providecommand \enquote  [1]{``#1''}%
\providecommand \bibnamefont  [1]{#1}%
\providecommand \bibfnamefont [1]{#1}%
\providecommand \citenamefont [1]{#1}%
\providecommand \href@noop [0]{\@secondoftwo}%
\providecommand \href [0]{\begingroup \@sanitize@url \@href}%
\providecommand \@href[1]{\@@startlink{#1}\@@href}%
\providecommand \@@href[1]{\endgroup#1\@@endlink}%
\providecommand \@sanitize@url [0]{\catcode `\\12\catcode `\$12\catcode
  `\&12\catcode `\#12\catcode `\^12\catcode `\_12\catcode `\%12\relax}%
\providecommand \@@startlink[1]{}%
\providecommand \@@endlink[0]{}%
\providecommand \url  [0]{\begingroup\@sanitize@url \@url }%
\providecommand \@url [1]{\endgroup\@href {#1}{\urlprefix }}%
\providecommand \urlprefix  [0]{URL }%
\providecommand \Eprint [0]{\href }%
\providecommand \doibase [0]{http://dx.doi.org/}%
\providecommand \selectlanguage [0]{\@gobble}%
\providecommand \bibinfo  [0]{\@secondoftwo}%
\providecommand \bibfield  [0]{\@secondoftwo}%
\providecommand \translation [1]{[#1]}%
\providecommand \BibitemOpen [0]{}%
\providecommand \bibitemStop [0]{}%
\providecommand \bibitemNoStop [0]{.\EOS\space}%
\providecommand \EOS [0]{\spacefactor3000\relax}%
\providecommand \BibitemShut  [1]{\csname bibitem#1\endcsname}%
\let\auto@bib@innerbib\@empty
%</preamble>
\end{thebibliography}%


\begin{thebibliography}{48}%
\makeatletter
\providecommand \@ifxundefined [1]{%
 \@ifx{#1\undefined}
}%
\providecommand \@ifnum [1]{%
 \ifnum #1\expandafter \@firstoftwo
 \else \expandafter \@secondoftwo
 \fi
}%
\providecommand \@ifx [1]{%
 \ifx #1\expandafter \@firstoftwo
 \else \expandafter \@secondoftwo
 \fi
}%
\providecommand \natexlab [1]{#1}%
\providecommand \enquote  [1]{``#1''}%
\providecommand \bibnamefont  [1]{#1}%
\providecommand \bibfnamefont [1]{#1}%
\providecommand \citenamefont [1]{#1}%
\providecommand \href@noop [0]{\@secondoftwo}%
\providecommand \href [0]{\begingroup \@sanitize@url \@href}%
\providecommand \@href[1]{\@@startlink{#1}\@@href}%
\providecommand \@@href[1]{\endgroup#1\@@endlink}%
\providecommand \@sanitize@url [0]{\catcode `\\12\catcode `\$12\catcode
  `\&12\catcode `\#12\catcode `\^12\catcode `\_12\catcode `\%12\relax}%
\providecommand \@@startlink[1]{}%
\providecommand \@@endlink[0]{}%
\providecommand \url  [0]{\begingroup\@sanitize@url \@url }%
\providecommand \@url [1]{\endgroup\@href {#1}{\urlprefix }}%
\providecommand \urlprefix  [0]{URL }%
\providecommand \Eprint [0]{\href }%
\providecommand \doibase [0]{http://dx.doi.org/}%
\providecommand \selectlanguage [0]{\@gobble}%
\providecommand \bibinfo  [0]{\@secondoftwo}%
\providecommand \bibfield  [0]{\@secondoftwo}%
\providecommand \translation [1]{[#1]}%
\providecommand \BibitemOpen [0]{}%
\providecommand \bibitemStop [0]{}%
\providecommand \bibitemNoStop [0]{.\EOS\space}%
\providecommand \EOS [0]{\spacefactor3000\relax}%
\providecommand \BibitemShut  [1]{\csname bibitem#1\endcsname}%
\let\auto@bib@innerbib\@empty
%</preamble>
\bibitem [{\citenamefont {Kittel}(2004)}]{kittel04}%
  \BibitemOpen
  \bibfield  {author} {\bibinfo {author} {\bibfnamefont {C.}~\bibnamefont
  {Kittel}},\ }\href@noop {} {\emph {\bibinfo {title} {Introduction to Solid
  State Physics}}},\ \bibinfo {edition} {8th}\ ed.\ (\bibinfo  {publisher}
  {Wiley},\ \bibinfo {year} {2004})\BibitemShut {NoStop}%
\bibitem [{\citenamefont {Mayergoyz}\ \emph {et~al.}(2009)\citenamefont
  {Mayergoyz}, \citenamefont {Bertotti},\ and\ \citenamefont
  {Serpico}}]{mayergoyz09}%
  \BibitemOpen
  \bibfield  {author} {\bibinfo {author} {\bibfnamefont {I.~D.}\ \bibnamefont
  {Mayergoyz}}, \bibinfo {author} {\bibfnamefont {G.}~\bibnamefont {Bertotti}},
  \ and\ \bibinfo {author} {\bibfnamefont {C.}~\bibnamefont {Serpico}},\
  }\href@noop {} {\emph {\bibinfo {title} {Nonlinear magnetization dynamics in
  nanosystems}}}\ (\bibinfo  {publisher} {Elsevier},\ \bibinfo {year}
  {2009})\BibitemShut {NoStop}%
\bibitem [{\citenamefont {Gurevich}\ and\ \citenamefont
  {Melkov}(1996)}]{gurevich96}%
  \BibitemOpen
  \bibfield  {author} {\bibinfo {author} {\bibfnamefont {A.~G.}\ \bibnamefont
  {Gurevich}}\ and\ \bibinfo {author} {\bibfnamefont {G.~A.}\ \bibnamefont
  {Melkov}},\ }\href@noop {} {\emph {\bibinfo {title} {Magnetization
  {O}scillations and {W}aves}}}\ (\bibinfo  {publisher} {CRC Press},\ \bibinfo
  {year} {1996})\BibitemShut {NoStop}%
\bibitem [{\citenamefont {Wigen}(1994)}]{wigen94}%
  \BibitemOpen
  \bibinfo {editor} {\bibfnamefont {P.~E.}\ \bibnamefont {Wigen}},\ ed.,\
  \href@noop {} {\emph {\bibinfo {title} {Nonlinear Phenomena and Chaos in
  Magnetic Materials}}}\ (\bibinfo  {publisher} {World ScientificWorld
  Scientific, Singapore},\ \bibinfo {year} {1994})\BibitemShut {NoStop}%
\bibitem [{\citenamefont {Mohseni}\ \emph {et~al.}(2013)\citenamefont
  {Mohseni}, \citenamefont {Sani}, \citenamefont {Persson}, \citenamefont
  {Nguyen}, \citenamefont {Chung}, \citenamefont {Pogoryelov}, \citenamefont
  {Muduli}, \citenamefont {Iacocca}, \citenamefont {Eklund}, \citenamefont
  {Dumas}, \citenamefont {Bonetti}, \citenamefont {Deac}, \citenamefont
  {Hoefer},\ and\ \citenamefont {Akerman}}]{mohseni13}%
  \BibitemOpen
  \bibfield  {author} {\bibinfo {author} {\bibfnamefont {S.~M.}\ \bibnamefont
  {Mohseni}}, \bibinfo {author} {\bibfnamefont {S.~R.}\ \bibnamefont {Sani}},
  \bibinfo {author} {\bibfnamefont {J.}~\bibnamefont {Persson}}, \bibinfo
  {author} {\bibfnamefont {T.~N.~Anh}\ \bibnamefont {Nguyen}}, \bibinfo
  {author} {\bibfnamefont {S.}~\bibnamefont {Chung}}, \bibinfo {author}
  {\bibfnamefont {Ye.}\ \bibnamefont {Pogoryelov}}, \bibinfo {author}
  {\bibfnamefont {P.~K.}\ \bibnamefont {Muduli}}, \bibinfo {author}
  {\bibfnamefont {E.}~\bibnamefont {Iacocca}}, \bibinfo {author} {\bibfnamefont
  {A.}~\bibnamefont {Eklund}}, \bibinfo {author} {\bibfnamefont {R.~K.}\
  \bibnamefont {Dumas}}, \bibinfo {author} {\bibfnamefont {S.}~\bibnamefont
  {Bonetti}}, \bibinfo {author} {\bibfnamefont {A.}~\bibnamefont {Deac}},
  \bibinfo {author} {\bibfnamefont {M.~A.}\ \bibnamefont {Hoefer}}, \ and\
  \bibinfo {author} {\bibfnamefont {J.}~\bibnamefont {Akerman}},\ }\bibfield
  {title} {\enquote {\bibinfo {title} {Spin torque-generated magnetic droplet
  solitons},}\ }\href {\doibase 10.1126/science.1230155} {\bibfield  {journal}
  {\bibinfo  {journal} {Science}\ }\textbf {\bibinfo {volume} {339}},\ \bibinfo
  {pages} {1295--1298} (\bibinfo {year} {2013})}\BibitemShut {NoStop}%
\bibitem [{\citenamefont {Rezende}\ and\ \citenamefont
  {de~Aguiar}(1990)}]{rezende90}%
  \BibitemOpen
  \bibfield  {author} {\bibinfo {author} {\bibfnamefont {S.~M.}\ \bibnamefont
  {Rezende}}\ and\ \bibinfo {author} {\bibfnamefont {F.~M.}\ \bibnamefont
  {de~Aguiar}},\ }\bibfield  {title} {\enquote {\bibinfo {title} {Spin-wave
  instabilities, auto-oscillations, and chaos in yttrium-iron-garnet},}\ }\href
  {\doibase 10.1109/5.56906} {\bibfield  {journal} {\bibinfo  {journal} {Proc.
  IEEE}\ }\textbf {\bibinfo {volume} {78}},\ \bibinfo {pages} {893--908}
  (\bibinfo {year} {1990})}\BibitemShut {NoStop}%
\bibitem [{\citenamefont {L'vov}(1994)}]{lvov94}%
  \BibitemOpen
  \bibfield  {author} {\bibinfo {author} {\bibfnamefont {Victor~S.}\
  \bibnamefont {L'vov}},\ }\href@noop {} {\emph {\bibinfo {title} {Wave
  Turbulence Under Parametric Excitation: Applications to Magnets}}}\ (\bibinfo
   {publisher} {Springer-Verlag},\ \bibinfo {year} {1994})\BibitemShut
  {NoStop}%
\bibitem [{\citenamefont {Petit-Watelot}\ \emph {et~al.}(2012)\citenamefont
  {Petit-Watelot}, \citenamefont {Kim}, \citenamefont {Ruotolo}, \citenamefont
  {Otxoa}, \citenamefont {Bouzehouane}, \citenamefont {Grollier}, \citenamefont
  {Vansteenkiste}, \citenamefont {de~Wiele}, \citenamefont {Cros},\ and\
  \citenamefont {Devolder}}]{petit-watelot12}%
  \BibitemOpen
  \bibfield  {author} {\bibinfo {author} {\bibfnamefont {S.}~\bibnamefont
  {Petit-Watelot}}, \bibinfo {author} {\bibfnamefont {J.-V.}\ \bibnamefont
  {Kim}}, \bibinfo {author} {\bibfnamefont {A.}~\bibnamefont {Ruotolo}},
  \bibinfo {author} {\bibfnamefont {R.~M.}\ \bibnamefont {Otxoa}}, \bibinfo
  {author} {\bibfnamefont {K.}~\bibnamefont {Bouzehouane}}, \bibinfo {author}
  {\bibfnamefont {J.}~\bibnamefont {Grollier}}, \bibinfo {author}
  {\bibfnamefont {A.}~\bibnamefont {Vansteenkiste}}, \bibinfo {author}
  {\bibfnamefont {B.~Van}\ \bibnamefont {de~Wiele}}, \bibinfo {author}
  {\bibfnamefont {V.}~\bibnamefont {Cros}}, \ and\ \bibinfo {author}
  {\bibfnamefont {T.}~\bibnamefont {Devolder}},\ }\bibfield  {title} {\enquote
  {\bibinfo {title} {Commensurability and chaos in magnetic vortex
  oscillations},}\ }\href {\doibase 10.1038/nphys2362} {\bibfield  {journal}
  {\bibinfo  {journal} {Nature Phys.}\ }\textbf {\bibinfo {volume} {8}},\
  \bibinfo {pages} {682--687} (\bibinfo {year} {2012})}\BibitemShut {NoStop}%
\bibitem [{\citenamefont {Demokritov}\ \emph {et~al.}(2006)\citenamefont
  {Demokritov}, \citenamefont {Demidov}, \citenamefont {Dzyapko}, \citenamefont
  {Melkov}, \citenamefont {Serga}, \citenamefont {Hillebrands},\ and\
  \citenamefont {Slavin}}]{demokritov06}%
  \BibitemOpen
  \bibfield  {author} {\bibinfo {author} {\bibfnamefont {S.~O.}\ \bibnamefont
  {Demokritov}}, \bibinfo {author} {\bibfnamefont {V.~E.}\ \bibnamefont
  {Demidov}}, \bibinfo {author} {\bibfnamefont {O.}~\bibnamefont {Dzyapko}},
  \bibinfo {author} {\bibfnamefont {G.~A.}\ \bibnamefont {Melkov}}, \bibinfo
  {author} {\bibfnamefont {A.~A.}\ \bibnamefont {Serga}}, \bibinfo {author}
  {\bibfnamefont {B.}~\bibnamefont {Hillebrands}}, \ and\ \bibinfo {author}
  {\bibfnamefont {A.~N.}\ \bibnamefont {Slavin}},\ }\bibfield  {title}
  {\enquote {\bibinfo {title} {Bose--{E}instein condensation of
  quasi-equilibrium magnons at room temperature under pumping},}\ }\href
  {\doibase 10.1038/nature05117} {\bibfield  {journal} {\bibinfo  {journal}
  {Nature}\ }\textbf {\bibinfo {volume} {443}},\ \bibinfo {pages} {430--433}
  (\bibinfo {year} {2006})}\BibitemShut {NoStop}%
\bibitem [{\citenamefont {Chumak}\ \emph {et~al.}(2015)\citenamefont {Chumak},
  \citenamefont {Vasyuchka}, \citenamefont {Serga},\ and\ \citenamefont
  {Hillebrands}}]{chumak15}%
  \BibitemOpen
  \bibfield  {author} {\bibinfo {author} {\bibfnamefont {A.~V.}\ \bibnamefont
  {Chumak}}, \bibinfo {author} {\bibfnamefont {V.~I.}\ \bibnamefont
  {Vasyuchka}}, \bibinfo {author} {\bibfnamefont {A.~A.}\ \bibnamefont
  {Serga}}, \ and\ \bibinfo {author} {\bibfnamefont {B.}~\bibnamefont
  {Hillebrands}},\ }\bibfield  {title} {\enquote {\bibinfo {title} {Magnon
  spintronics},}\ }\href {\doibase 10.1038/nphys3347} {\bibfield  {journal}
  {\bibinfo  {journal} {Nature Phys.}\ }\textbf {\bibinfo {volume} {11}},\
  \bibinfo {pages} {453--461} (\bibinfo {year} {2015})}\BibitemShut {NoStop}%
\bibitem [{\citenamefont {Kiselev}\ \emph {et~al.}(2003)\citenamefont
  {Kiselev}, \citenamefont {Sankey}, \citenamefont {Krivorotov}, \citenamefont
  {Emley}, \citenamefont {Schoelkopf}, \citenamefont {Buhrman},\ and\
  \citenamefont {Ralph}}]{kiselev03}%
  \BibitemOpen
  \bibfield  {author} {\bibinfo {author} {\bibfnamefont {S.~I.}\ \bibnamefont
  {Kiselev}}, \bibinfo {author} {\bibfnamefont {J.~C.}\ \bibnamefont {Sankey}},
  \bibinfo {author} {\bibfnamefont {I.~N.}\ \bibnamefont {Krivorotov}},
  \bibinfo {author} {\bibfnamefont {N.~C.}\ \bibnamefont {Emley}}, \bibinfo
  {author} {\bibfnamefont {R.~J.}\ \bibnamefont {Schoelkopf}}, \bibinfo
  {author} {\bibfnamefont {R.~A.}\ \bibnamefont {Buhrman}}, \ and\ \bibinfo
  {author} {\bibfnamefont {D.~C.}\ \bibnamefont {Ralph}},\ }\bibfield  {title}
  {\enquote {\bibinfo {title} {Microwave oscillations of a nanomagnet driven by
  a spin-polarized current},}\ }\href {\doibase 10.1038/nature01967} {\bibfield
   {journal} {\bibinfo  {journal} {Nature}\ }\textbf {\bibinfo {volume}
  {425}},\ \bibinfo {pages} {380--383} (\bibinfo {year} {2003})}\BibitemShut
  {NoStop}%
\bibitem [{\citenamefont {Houssameddine}\ \emph {et~al.}(2007)\citenamefont
  {Houssameddine}, \citenamefont {Ebels}, \citenamefont {Delaët}, \citenamefont
  {Rodmacq}, \citenamefont {Firastrau}, \citenamefont {Ponthenier},
  \citenamefont {Brunet}, \citenamefont {Thirion}, \citenamefont {Michel},
  \citenamefont {Prejbeanu-Buda}, \citenamefont {Cyrille}, \citenamefont
  {Redon},\ and\ \citenamefont {Dieny}}]{houssameddine07}%
  \BibitemOpen
  \bibfield  {author} {\bibinfo {author} {\bibfnamefont {D.}~\bibnamefont
  {Houssameddine}}, \bibinfo {author} {\bibfnamefont {U.}~\bibnamefont
  {Ebels}}, \bibinfo {author} {\bibfnamefont {B.}~\bibnamefont {Delaët}},
  \bibinfo {author} {\bibfnamefont {B.}~\bibnamefont {Rodmacq}}, \bibinfo
  {author} {\bibfnamefont {I.}~\bibnamefont {Firastrau}}, \bibinfo {author}
  {\bibfnamefont {F.}~\bibnamefont {Ponthenier}}, \bibinfo {author}
  {\bibfnamefont {M.}~\bibnamefont {Brunet}}, \bibinfo {author} {\bibfnamefont
  {C.}~\bibnamefont {Thirion}}, \bibinfo {author} {\bibfnamefont {J.-P.}\
  \bibnamefont {Michel}}, \bibinfo {author} {\bibfnamefont {L.}~\bibnamefont
  {Prejbeanu-Buda}}, \bibinfo {author} {\bibfnamefont {M.-C.}\ \bibnamefont
  {Cyrille}}, \bibinfo {author} {\bibfnamefont {O.}~\bibnamefont {Redon}}, \
  and\ \bibinfo {author} {\bibfnamefont {B.}~\bibnamefont {Dieny}},\ }\bibfield
   {title} {\enquote {\bibinfo {title} {Spin-torque oscillator using a
  perpendicular polarizer and a planar free layer},}\ }\href {\doibase
  10.1038/nmat1905} {\bibfield  {journal} {\bibinfo  {journal} {Nature Mater.}\
  }\textbf {\bibinfo {volume} {6}},\ \bibinfo {pages} {447--453} (\bibinfo
  {year} {2007})}\BibitemShut {NoStop}%
\bibitem [{\citenamefont {Chen}\ \emph {et~al.}(2009)\citenamefont {Chen},
  \citenamefont {de~Loubens}, \citenamefont {Beaujour}, \citenamefont {Sun},\
  and\ \citenamefont {Kent}}]{chen09}%
  \BibitemOpen
  \bibfield  {author} {\bibinfo {author} {\bibfnamefont {W.}~\bibnamefont
  {Chen}}, \bibinfo {author} {\bibfnamefont {G.}~\bibnamefont {de~Loubens}},
  \bibinfo {author} {\bibfnamefont {J.-M.~L.}\ \bibnamefont {Beaujour}},
  \bibinfo {author} {\bibfnamefont {J.~Z.}\ \bibnamefont {Sun}}, \ and\
  \bibinfo {author} {\bibfnamefont {A.~D.}\ \bibnamefont {Kent}},\ }\bibfield
  {title} {\enquote {\bibinfo {title} {Spin-torque driven ferromagnetic
  resonance in a nonlinear regime},}\ }\href {\doibase 10.1063/1.3254242}
  {\bibfield  {journal} {\bibinfo  {journal} {Appl. Phys. Lett.}\ }\textbf
  {\bibinfo {volume} {95}},\ \bibinfo {eid} {172513} (\bibinfo {year}
  {2009})}\BibitemShut {NoStop}%
\bibitem [{\citenamefont {Hamadeh}\ \emph {et~al.}(2012)\citenamefont
  {Hamadeh}, \citenamefont {de~Loubens}, \citenamefont {Naletov}, \citenamefont
  {Grollier}, \citenamefont {Ulysse}, \citenamefont {Cros},\ and\ \citenamefont
  {Klein}}]{hamadeh12}%
  \BibitemOpen
  \bibfield  {author} {\bibinfo {author} {\bibfnamefont {A.}~\bibnamefont
  {Hamadeh}}, \bibinfo {author} {\bibfnamefont {G.}~\bibnamefont {de~Loubens}},
  \bibinfo {author} {\bibfnamefont {V.~V.}\ \bibnamefont {Naletov}}, \bibinfo
  {author} {\bibfnamefont {J.}~\bibnamefont {Grollier}}, \bibinfo {author}
  {\bibfnamefont {C.}~\bibnamefont {Ulysse}}, \bibinfo {author} {\bibfnamefont
  {V.}~\bibnamefont {Cros}}, \ and\ \bibinfo {author} {\bibfnamefont
  {O.}~\bibnamefont {Klein}},\ }\bibfield  {title} {\enquote {\bibinfo {title}
  {Autonomous and forced dynamics in a spin-transfer nano-oscillator:
  Quantitative magnetic-resonance force microscopy},}\ }\href {\doibase
  10.1103/PhysRevB.85.140408} {\bibfield  {journal} {\bibinfo  {journal} {Phys.
  Rev. B}\ }\textbf {\bibinfo {volume} {85}},\ \bibinfo {pages} {140408}
  (\bibinfo {year} {2012})}\BibitemShut {NoStop}%
\bibitem [{\citenamefont {Collet}\ \emph {et~al.}(2016)\citenamefont {Collet},
  \citenamefont {de~Milly}, \citenamefont {d'Allivy Kelly}, \citenamefont
  {Naletov}, \citenamefont {Bernard}, \citenamefont {Bortolotti}, \citenamefont
  {{Ben Youssef}}, \citenamefont {Demidov}, \citenamefont {Demokritov},
  \citenamefont {Prieto}, \citenamefont {Mu{\~n}oz}, \citenamefont {Cros},
  \citenamefont {Anane}, \citenamefont {de~Loubens},\ and\ \citenamefont
  {Klein}}]{collet16}%
  \BibitemOpen
  \bibfield  {author} {\bibinfo {author} {\bibfnamefont {M.}~\bibnamefont
  {Collet}}, \bibinfo {author} {\bibfnamefont {X.}~\bibnamefont {de~Milly}},
  \bibinfo {author} {\bibfnamefont {O.}~\bibnamefont {d'Allivy Kelly}},
  \bibinfo {author} {\bibfnamefont {V.V.}\ \bibnamefont {Naletov}}, \bibinfo
  {author} {\bibfnamefont {R.}~\bibnamefont {Bernard}}, \bibinfo {author}
  {\bibfnamefont {P.}~\bibnamefont {Bortolotti}}, \bibinfo {author}
  {\bibfnamefont {J.}~\bibnamefont {{Ben Youssef}}}, \bibinfo {author}
  {\bibfnamefont {V.E.}\ \bibnamefont {Demidov}}, \bibinfo {author}
  {\bibfnamefont {S.O.}\ \bibnamefont {Demokritov}}, \bibinfo {author}
  {\bibfnamefont {J.L.}\ \bibnamefont {Prieto}}, \bibinfo {author}
  {\bibfnamefont {M.}~\bibnamefont {Mu{\~n}oz}}, \bibinfo {author}
  {\bibfnamefont {V.}~\bibnamefont {Cros}}, \bibinfo {author} {\bibfnamefont
  {A.}~\bibnamefont {Anane}}, \bibinfo {author} {\bibfnamefont
  {G.}~\bibnamefont {de~Loubens}}, \ and\ \bibinfo {author} {\bibfnamefont
  {O.}~\bibnamefont {Klein}},\ }\bibfield  {title} {\enquote {\bibinfo {title}
  {Generation of coherent spin-wave modes in yttrium iron garnet microdiscs by
  spin-orbit torque},}\ }\href {\doibase 10.1038/ncomms10377} {\bibfield
  {journal} {\bibinfo  {journal} {Nature Commun.}\ }\textbf {\bibinfo {volume}
  {7}},\ \bibinfo {pages} {10377} (\bibinfo {year} {2016})}\BibitemShut
  {NoStop}%
\bibitem [{\citenamefont {Slavin}\ and\ \citenamefont
  {Tiberkevich}(2009)}]{slavin09}%
  \BibitemOpen
  \bibfield  {author} {\bibinfo {author} {\bibfnamefont {A.}~\bibnamefont
  {Slavin}}\ and\ \bibinfo {author} {\bibfnamefont {V.}~\bibnamefont
  {Tiberkevich}},\ }\bibfield  {title} {\enquote {\bibinfo {title} {Nonlinear
  auto-oscillator theory of microwave generation by spin-polarized current},}\
  }\href {\doibase 10.1109/TMAG.2008.2009935} {\bibfield  {journal} {\bibinfo
  {journal} {IEEE Trans. Magn.}\ }\textbf {\bibinfo {volume} {45}},\ \bibinfo
  {pages} {1875--1918} (\bibinfo {year} {2009})}\BibitemShut {NoStop}%
\bibitem [{\citenamefont {Torrejon}\ \emph {et~al.}(2017)\citenamefont
  {Torrejon}, \citenamefont {Riou}, \citenamefont {Araujo}, \citenamefont
  {Tsunegi}, \citenamefont {Khalsa}, \citenamefont {Querlioz}, \citenamefont
  {Bortolotti}, \citenamefont {Cros}, \citenamefont {Yakushiji}, \citenamefont
  {Fukushima}, \citenamefont {Kubota}, \citenamefont {Yuasa}, \citenamefont
  {Stiles},\ and\ \citenamefont {Grollier}}]{torrejon17}%
  \BibitemOpen
  \bibfield  {author} {\bibinfo {author} {\bibfnamefont {J.}~\bibnamefont
  {Torrejon}}, \bibinfo {author} {\bibfnamefont {M.}~\bibnamefont {Riou}},
  \bibinfo {author} {\bibfnamefont {F.~Abreu}\ \bibnamefont {Araujo}}, \bibinfo
  {author} {\bibfnamefont {S.}~\bibnamefont {Tsunegi}}, \bibinfo {author}
  {\bibfnamefont {G.}~\bibnamefont {Khalsa}}, \bibinfo {author} {\bibfnamefont
  {D.}~\bibnamefont {Querlioz}}, \bibinfo {author} {\bibfnamefont
  {P.}~\bibnamefont {Bortolotti}}, \bibinfo {author} {\bibfnamefont
  {V.}~\bibnamefont {Cros}}, \bibinfo {author} {\bibfnamefont {K.}~\bibnamefont
  {Yakushiji}}, \bibinfo {author} {\bibfnamefont {A.}~\bibnamefont
  {Fukushima}}, \bibinfo {author} {\bibfnamefont {H.}~\bibnamefont {Kubota}},
  \bibinfo {author} {\bibfnamefont {S.}~\bibnamefont {Yuasa}}, \bibinfo
  {author} {\bibfnamefont {M.~D.}\ \bibnamefont {Stiles}}, \ and\ \bibinfo
  {author} {\bibfnamefont {J.}~\bibnamefont {Grollier}},\ }\bibfield  {title}
  {\enquote {\bibinfo {title} {Neuromorphic computing with nanoscale spintronic
  oscillators},}\ }\href {\doibase 10.1038/nature23011} {\bibfield  {journal}
  {\bibinfo  {journal} {Nature}\ }\textbf {\bibinfo {volume} {547}},\ \bibinfo
  {pages} {428--431} (\bibinfo {year} {2017})}\BibitemShut {NoStop}%
\bibitem [{\citenamefont {Romera}\ \emph {et~al.}(2018)\citenamefont {Romera},
  \citenamefont {Talatchian}, \citenamefont {Tsunegi}, \citenamefont
  {Abreu~Araujo}, \citenamefont {Cros}, \citenamefont {Bortolotti},
  \citenamefont {Trastoy}, \citenamefont {Yakushiji}, \citenamefont
  {Fukushima}, \citenamefont {Kubota} \emph {et~al.}}]{romera18}%
  \BibitemOpen
  \bibfield  {author} {\bibinfo {author} {\bibfnamefont {M.}~\bibnamefont
  {Romera}}, \bibinfo {author} {\bibfnamefont {P.}~\bibnamefont {Talatchian}},
  \bibinfo {author} {\bibfnamefont {S.}~\bibnamefont {Tsunegi}}, \bibinfo
  {author} {\bibfnamefont {F.}~\bibnamefont {Abreu~Araujo}}, \bibinfo {author}
  {\bibfnamefont {V.}~\bibnamefont {Cros}}, \bibinfo {author} {\bibfnamefont
  {P.}~\bibnamefont {Bortolotti}}, \bibinfo {author} {\bibfnamefont
  {J.}~\bibnamefont {Trastoy}}, \bibinfo {author} {\bibfnamefont
  {K.}~\bibnamefont {Yakushiji}}, \bibinfo {author} {\bibfnamefont
  {A.}~\bibnamefont {Fukushima}}, \bibinfo {author} {\bibfnamefont
  {H.}~\bibnamefont {Kubota}},  \emph {et~al.},\ }\bibfield  {title} {\enquote
  {\bibinfo {title} {Vowel recognition with four coupled spin-torque
  nano-oscillators},}\ }\href {\doibase 10.1038/s41586-018-0632-y} {\bibfield
  {journal} {\bibinfo  {journal} {Nature}\ }\textbf {\bibinfo {volume} {563}},\
  \bibinfo {pages} {230--234} (\bibinfo {year} {2018})}\BibitemShut {NoStop}%
\bibitem [{\citenamefont {Suhl}(1957)}]{suhl57}%
  \BibitemOpen
  \bibfield  {author} {\bibinfo {author} {\bibfnamefont {H.}~\bibnamefont
  {Suhl}},\ }\bibfield  {title} {\enquote {\bibinfo {title} {The theory of
  ferromagnetic resonance at high signal powers},}\ }\href {\doibase
  10.1016/0022-3697(57)90010-0} {\bibfield  {journal} {\bibinfo  {journal} {J.
  {P}hys. {C}hem. {S}olids}\ }\textbf {\bibinfo {volume} {1}},\ \bibinfo
  {pages} {209--227} (\bibinfo {year} {1957})}\BibitemShut {NoStop}%
\bibitem [{\citenamefont {Anderson}\ and\ \citenamefont
  {Suhl}(1955)}]{anderson55}%
  \BibitemOpen
  \bibfield  {author} {\bibinfo {author} {\bibfnamefont {P.~W.}\ \bibnamefont
  {Anderson}}\ and\ \bibinfo {author} {\bibfnamefont {H.}~\bibnamefont
  {Suhl}},\ }\bibfield  {title} {\enquote {\bibinfo {title} {Instability in the
  motion of ferromagnets at high microwave power levels},}\ }\href {\doibase
  10.1103/PhysRev.100.1788} {\bibfield  {journal} {\bibinfo  {journal} {Phys.
  Rev.}\ }\textbf {\bibinfo {volume} {100}},\ \bibinfo {pages} {1788--1789}
  (\bibinfo {year} {1955})}\BibitemShut {NoStop}%
\bibitem [{\citenamefont {Kobljanskyj}\ \emph {et~al.}(2012)\citenamefont
  {Kobljanskyj}, \citenamefont {Melkov}, \citenamefont {Guslienko},
  \citenamefont {Novosad}, \citenamefont {Bader}, \citenamefont {Kostylev},\
  and\ \citenamefont {Slavin}}]{kobljanskyj12}%
  \BibitemOpen
  \bibfield  {author} {\bibinfo {author} {\bibfnamefont {Y.}~\bibnamefont
  {Kobljanskyj}}, \bibinfo {author} {\bibfnamefont {G.}~\bibnamefont {Melkov}},
  \bibinfo {author} {\bibfnamefont {K.}~\bibnamefont {Guslienko}}, \bibinfo
  {author} {\bibfnamefont {V.}~\bibnamefont {Novosad}}, \bibinfo {author}
  {\bibfnamefont {S.~D.}\ \bibnamefont {Bader}}, \bibinfo {author}
  {\bibfnamefont {M.}~\bibnamefont {Kostylev}}, \ and\ \bibinfo {author}
  {\bibfnamefont {A.}~\bibnamefont {Slavin}},\ }\bibfield  {title} {\enquote
  {\bibinfo {title} {Nano-structured magnetic metamaterial with enhanced
  nonlinear properties},}\ }\href {\doibase 10.1038/srep00478} {\bibfield
  {journal} {\bibinfo  {journal} {Sci. Rep.}\ }\textbf {\bibinfo {volume}
  {2}},\ \bibinfo {pages} {478} (\bibinfo {year} {2012})}\BibitemShut {NoStop}%
\bibitem [{\citenamefont {Melkov}\ \emph {et~al.}(2013)\citenamefont {Melkov},
  \citenamefont {Slobodianiuk}, \citenamefont {Tiberkevich}, \citenamefont
  {de~Loubens}, \citenamefont {Klein},\ and\ \citenamefont
  {Slavin}}]{melkov13}%
  \BibitemOpen
  \bibfield  {author} {\bibinfo {author} {\bibfnamefont {G.A.}\ \bibnamefont
  {Melkov}}, \bibinfo {author} {\bibfnamefont {D.V.}\ \bibnamefont
  {Slobodianiuk}}, \bibinfo {author} {\bibfnamefont {V.S.}\ \bibnamefont
  {Tiberkevich}}, \bibinfo {author} {\bibfnamefont {G.}~\bibnamefont
  {de~Loubens}}, \bibinfo {author} {\bibfnamefont {O.}~\bibnamefont {Klein}}, \
  and\ \bibinfo {author} {\bibfnamefont {A.N.}\ \bibnamefont {Slavin}},\
  }\bibfield  {title} {\enquote {\bibinfo {title} {Nonlinear ferromagnetic
  resonance in nanostructures having discrete spectrum of spin-wave modes},}\
  }\href {\doibase 10.1109/LMAG.2013.2278682} {\bibfield  {journal} {\bibinfo
  {journal} {IEEE Magn. Lett.}\ }\textbf {\bibinfo {volume} {4}},\ \bibinfo
  {pages} {4000504} (\bibinfo {year} {2013})}\BibitemShut {NoStop}%
\bibitem [{\citenamefont {McMichael}\ and\ \citenamefont
  {Wigen}(1990)}]{mcmichael90}%
  \BibitemOpen
  \bibfield  {author} {\bibinfo {author} {\bibfnamefont {R.~D.}\ \bibnamefont
  {McMichael}}\ and\ \bibinfo {author} {\bibfnamefont {P.~E.}\ \bibnamefont
  {Wigen}},\ }\bibfield  {title} {\enquote {\bibinfo {title} {High-power
  ferromagnetic resonance without a degenerate spin-wave manifold},}\ }\href
  {\doibase 10.1103/PhysRevLett.64.64} {\bibfield  {journal} {\bibinfo
  {journal} {Phys. Rev. Lett.}\ }\textbf {\bibinfo {volume} {64}},\ \bibinfo
  {pages} {64--67} (\bibinfo {year} {1990})}\BibitemShut {NoStop}%
\bibitem [{\citenamefont {Gulyaev}\ \emph {et~al.}(2000)\citenamefont
  {Gulyaev}, \citenamefont {Zil'berman}, \citenamefont {Temiryazev},\ and\
  \citenamefont {Tikhomirova}}]{gulyaev00}%
  \BibitemOpen
  \bibfield  {author} {\bibinfo {author} {\bibfnamefont {Yu.~V.}\ \bibnamefont
  {Gulyaev}}, \bibinfo {author} {\bibfnamefont {P.~E.}\ \bibnamefont
  {Zil'berman}}, \bibinfo {author} {\bibfnamefont {A.~G.}\ \bibnamefont
  {Temiryazev}}, \ and\ \bibinfo {author} {\bibfnamefont {M.~P.}\ \bibnamefont
  {Tikhomirova}},\ }\bibfield  {title} {\enquote {\bibinfo {title} {Principal
  mode of the nonlinear spin-wave resonance in perpendicular magnetized ferrite
  films},}\ }\href {\doibase 10.1134/1.1131354} {\bibfield  {journal} {\bibinfo
   {journal} {Phys. Solid State}\ }\textbf {\bibinfo {volume} {42}},\ \bibinfo
  {pages} {1094--1099} (\bibinfo {year} {2000})}\BibitemShut {NoStop}%
\bibitem [{\citenamefont {Gnatzig}\ \emph {et~al.}(1987)\citenamefont
  {Gnatzig}, \citenamefont {D\"otsch}, \citenamefont {Ye},\ and\ \citenamefont
  {Brockmeyer}}]{gnatzig87}%
  \BibitemOpen
  \bibfield  {author} {\bibinfo {author} {\bibfnamefont {K.}~\bibnamefont
  {Gnatzig}}, \bibinfo {author} {\bibfnamefont {H.}~\bibnamefont {D\"otsch}},
  \bibinfo {author} {\bibfnamefont {M.}~\bibnamefont {Ye}}, \ and\ \bibinfo
  {author} {\bibfnamefont {A.}~\bibnamefont {Brockmeyer}},\ }\bibfield  {title}
  {\enquote {\bibinfo {title} {Ferrimagnetic resonance in garnet films at large
  precession angles},}\ }\href {\doibase 10.1063/1.338988} {\bibfield
  {journal} {\bibinfo  {journal} {J. Appl. Phys.}\ }\textbf {\bibinfo {volume}
  {62}},\ \bibinfo {pages} {4839--4843} (\bibinfo {year} {1987})}\BibitemShut
  {NoStop}%
\bibitem [{\citenamefont {Bertotti}\ \emph
  {et~al.}(2001{\natexlab{a}})\citenamefont {Bertotti}, \citenamefont
  {Serpico},\ and\ \citenamefont {Mayergoyz}}]{bertotti01}%
  \BibitemOpen
  \bibfield  {author} {\bibinfo {author} {\bibfnamefont {G.}~\bibnamefont
  {Bertotti}}, \bibinfo {author} {\bibfnamefont {C.}~\bibnamefont {Serpico}}, \
  and\ \bibinfo {author} {\bibfnamefont {I.~D.}\ \bibnamefont {Mayergoyz}},\
  }\bibfield  {title} {\enquote {\bibinfo {title} {Nonlinear magnetization
  dynamics under circularly polarized field},}\ }\href {\doibase
  10.1103/PhysRevLett.86.724} {\bibfield  {journal} {\bibinfo  {journal} {Phys.
  Rev. Lett.}\ }\textbf {\bibinfo {volume} {86}},\ \bibinfo {pages} {724--727}
  (\bibinfo {year} {2001}{\natexlab{a}})}\BibitemShut {NoStop}%
\bibitem [{\citenamefont {Bertotti}\ \emph
  {et~al.}(2001{\natexlab{b}})\citenamefont {Bertotti}, \citenamefont
  {Mayergoyz},\ and\ \citenamefont {Serpico}}]{bertotti01a}%
  \BibitemOpen
  \bibfield  {author} {\bibinfo {author} {\bibfnamefont {G.}~\bibnamefont
  {Bertotti}}, \bibinfo {author} {\bibfnamefont {I.~D.}\ \bibnamefont
  {Mayergoyz}}, \ and\ \bibinfo {author} {\bibfnamefont {C.}~\bibnamefont
  {Serpico}},\ }\bibfield  {title} {\enquote {\bibinfo {title} {Spin-wave
  instabilities in large-scale nonlinear magnetization dynamics},}\ }\href
  {\doibase 10.1103/PhysRevLett.87.217203} {\bibfield  {journal} {\bibinfo
  {journal} {Phys. Rev. Lett.}\ }\textbf {\bibinfo {volume} {87}},\ \bibinfo
  {pages} {217203} (\bibinfo {year} {2001}{\natexlab{b}})}\BibitemShut
  {NoStop}%
\bibitem [{\citenamefont {Fetisov}\ \emph {et~al.}(1999)\citenamefont
  {Fetisov}, \citenamefont {Patton},\ and\ \citenamefont
  {Synogach}}]{fetisov99}%
  \BibitemOpen
  \bibfield  {author} {\bibinfo {author} {\bibfnamefont {Y.K.}\ \bibnamefont
  {Fetisov}}, \bibinfo {author} {\bibfnamefont {Carl~E.}\ \bibnamefont
  {Patton}}, \ and\ \bibinfo {author} {\bibfnamefont {V.T.}\ \bibnamefont
  {Synogach}},\ }\bibfield  {title} {\enquote {\bibinfo {title} {Nonlinear
  ferromagnetic resonance and foldover in yttrium iron garnet thin films --
  inadequacy of the classical model},}\ }\href {\doibase 10.1109/20.809144}
  {\bibfield  {journal} {\bibinfo  {journal} {IEEE Trans. Magn.}\ }\textbf
  {\bibinfo {volume} {35}},\ \bibinfo {pages} {4511--4521} (\bibinfo {year}
  {1999})}\BibitemShut {NoStop}%
\bibitem [{\citenamefont {Gui}\ \emph {et~al.}(2009)\citenamefont {Gui},
  \citenamefont {Wirthmann},\ and\ \citenamefont {Hu}}]{gui09a}%
  \BibitemOpen
  \bibfield  {author} {\bibinfo {author} {\bibfnamefont {Y.~S.}\ \bibnamefont
  {Gui}}, \bibinfo {author} {\bibfnamefont {A.}~\bibnamefont {Wirthmann}}, \
  and\ \bibinfo {author} {\bibfnamefont {C.-M.}\ \bibnamefont {Hu}},\
  }\bibfield  {title} {\enquote {\bibinfo {title} {Foldover ferromagnetic
  resonance and damping in permalloy microstrips},}\ }\href {\doibase
  10.1103/PhysRevB.80.184422} {\bibfield  {journal} {\bibinfo  {journal} {Phys.
  Rev. B}\ }\textbf {\bibinfo {volume} {80}},\ \bibinfo {pages} {184422}
  (\bibinfo {year} {2009})}\BibitemShut {NoStop}%
\bibitem [{\citenamefont {Serga}\ \emph {et~al.}(2010)\citenamefont {Serga},
  \citenamefont {Chumak},\ and\ \citenamefont {Hillebrands}}]{serga10}%
  \BibitemOpen
  \bibfield  {author} {\bibinfo {author} {\bibfnamefont {A.~A.}\ \bibnamefont
  {Serga}}, \bibinfo {author} {\bibfnamefont {A.~V.}\ \bibnamefont {Chumak}}, \
  and\ \bibinfo {author} {\bibfnamefont {B.}~\bibnamefont {Hillebrands}},\
  }\bibfield  {title} {\enquote {\bibinfo {title} {{YIG} magnonics},}\ }\href
  {\doibase 10.1088/0022-3727/43/26/264002} {\bibfield  {journal} {\bibinfo
  {journal} {J. Phys. D: Appl. Phys.}\ }\textbf {\bibinfo {volume} {43}},\
  \bibinfo {pages} {264002} (\bibinfo {year} {2010})}\BibitemShut {NoStop}%
\bibitem [{\citenamefont {d'Allivy Kelly}\ \emph {et~al.}(2013)\citenamefont
  {d'Allivy Kelly}, \citenamefont {Anane}, \citenamefont {Bernard},
  \citenamefont {{Ben Youssef}}, \citenamefont {Hahn}, \citenamefont
  {Molpeceres}, \citenamefont {Carretero}, \citenamefont {Jacquet},
  \citenamefont {Deranlot}, \citenamefont {Bortolotti}, \citenamefont
  {Lebourgeois}, \citenamefont {Mage}, \citenamefont {de~Loubens},
  \citenamefont {Klein}, \citenamefont {Cros},\ and\ \citenamefont
  {Fert}}]{kelly13}%
  \BibitemOpen
  \bibfield  {author} {\bibinfo {author} {\bibfnamefont {O.}~\bibnamefont
  {d'Allivy Kelly}}, \bibinfo {author} {\bibfnamefont {A.}~\bibnamefont
  {Anane}}, \bibinfo {author} {\bibfnamefont {R.}~\bibnamefont {Bernard}},
  \bibinfo {author} {\bibfnamefont {J.}~\bibnamefont {{Ben Youssef}}}, \bibinfo
  {author} {\bibfnamefont {C.}~\bibnamefont {Hahn}}, \bibinfo {author}
  {\bibfnamefont {A~H.}\ \bibnamefont {Molpeceres}}, \bibinfo {author}
  {\bibfnamefont {C.}~\bibnamefont {Carretero}}, \bibinfo {author}
  {\bibfnamefont {E.}~\bibnamefont {Jacquet}}, \bibinfo {author} {\bibfnamefont
  {C.}~\bibnamefont {Deranlot}}, \bibinfo {author} {\bibfnamefont
  {P.}~\bibnamefont {Bortolotti}}, \bibinfo {author} {\bibfnamefont
  {R.}~\bibnamefont {Lebourgeois}}, \bibinfo {author} {\bibfnamefont {J.-C.}\
  \bibnamefont {Mage}}, \bibinfo {author} {\bibfnamefont {G.}~\bibnamefont
  {de~Loubens}}, \bibinfo {author} {\bibfnamefont {O.}~\bibnamefont {Klein}},
  \bibinfo {author} {\bibfnamefont {V.}~\bibnamefont {Cros}}, \ and\ \bibinfo
  {author} {\bibfnamefont {A.}~\bibnamefont {Fert}},\ }\bibfield  {title}
  {\enquote {\bibinfo {title} {Inverse spin hall effect in nanometer-thick
  yttrium iron garnet/{Pt} system},}\ }\href {\doibase 10.1063/1.4819157}
  {\bibfield  {journal} {\bibinfo  {journal} {Appl. Phys. Lett.}\ }\textbf
  {\bibinfo {volume} {103}},\ \bibinfo {pages} {082408} (\bibinfo {year}
  {2013})}\BibitemShut {NoStop}%
\bibitem [{\citenamefont {Klein}\ \emph {et~al.}(2008)\citenamefont {Klein},
  \citenamefont {de~Loubens}, \citenamefont {Naletov}, \citenamefont {Boust},
  \citenamefont {Guillet}, \citenamefont {Hurdequint}, \citenamefont
  {Leksikov}, \citenamefont {Slavin}, \citenamefont {Tiberkevich},\ and\
  \citenamefont {Vukadinovic}}]{klein08}%
  \BibitemOpen
  \bibfield  {author} {\bibinfo {author} {\bibfnamefont {O.}~\bibnamefont
  {Klein}}, \bibinfo {author} {\bibfnamefont {G.}~\bibnamefont {de~Loubens}},
  \bibinfo {author} {\bibfnamefont {V.~V.}\ \bibnamefont {Naletov}}, \bibinfo
  {author} {\bibfnamefont {F.}~\bibnamefont {Boust}}, \bibinfo {author}
  {\bibfnamefont {T.}~\bibnamefont {Guillet}}, \bibinfo {author} {\bibfnamefont
  {H.}~\bibnamefont {Hurdequint}}, \bibinfo {author} {\bibfnamefont
  {A.}~\bibnamefont {Leksikov}}, \bibinfo {author} {\bibfnamefont {A.~N.}\
  \bibnamefont {Slavin}}, \bibinfo {author} {\bibfnamefont {V.~S.}\
  \bibnamefont {Tiberkevich}}, \ and\ \bibinfo {author} {\bibfnamefont
  {N.}~\bibnamefont {Vukadinovic}},\ }\bibfield  {title} {\enquote {\bibinfo
  {title} {Ferromagnetic resonance force spectroscopy of individual
  submicron-size samples},}\ }\href {\doibase 10.1103/PhysRevB.78.144410}
  {\bibfield  {journal} {\bibinfo  {journal} {Phys. Rev. B}\ }\textbf {\bibinfo
  {volume} {78}},\ \bibinfo {eid} {144410} (\bibinfo {year}
  {2008})}\BibitemShut {NoStop}%
\bibitem [{\citenamefont {Naletov}\ \emph {et~al.}(2011)\citenamefont
  {Naletov}, \citenamefont {de~Loubens}, \citenamefont {Albuquerque},
  \citenamefont {Borlenghi}, \citenamefont {Cros}, \citenamefont {Faini},
  \citenamefont {Grollier}, \citenamefont {Hurdequint}, \citenamefont
  {Locatelli}, \citenamefont {Pigeau}, \citenamefont {Slavin}, \citenamefont
  {Tiberkevich}, \citenamefont {Ulysse}, \citenamefont {Valet},\ and\
  \citenamefont {Klein}}]{naletov11}%
  \BibitemOpen
  \bibfield  {author} {\bibinfo {author} {\bibfnamefont {V.~V.}\ \bibnamefont
  {Naletov}}, \bibinfo {author} {\bibfnamefont {G.}~\bibnamefont {de~Loubens}},
  \bibinfo {author} {\bibfnamefont {G.}~\bibnamefont {Albuquerque}}, \bibinfo
  {author} {\bibfnamefont {S.}~\bibnamefont {Borlenghi}}, \bibinfo {author}
  {\bibfnamefont {V.}~\bibnamefont {Cros}}, \bibinfo {author} {\bibfnamefont
  {G.}~\bibnamefont {Faini}}, \bibinfo {author} {\bibfnamefont
  {J.}~\bibnamefont {Grollier}}, \bibinfo {author} {\bibfnamefont
  {H.}~\bibnamefont {Hurdequint}}, \bibinfo {author} {\bibfnamefont
  {N.}~\bibnamefont {Locatelli}}, \bibinfo {author} {\bibfnamefont
  {B.}~\bibnamefont {Pigeau}}, \bibinfo {author} {\bibfnamefont {A.~N.}\
  \bibnamefont {Slavin}}, \bibinfo {author} {\bibfnamefont {V.~S.}\
  \bibnamefont {Tiberkevich}}, \bibinfo {author} {\bibfnamefont
  {C.}~\bibnamefont {Ulysse}}, \bibinfo {author} {\bibfnamefont
  {T.}~\bibnamefont {Valet}}, \ and\ \bibinfo {author} {\bibfnamefont
  {O.}~\bibnamefont {Klein}},\ }\bibfield  {title} {\enquote {\bibinfo {title}
  {Identification and selection rules of the spin-wave eigenmodes in a normally
  magnetized nanopillar},}\ }\href {\doibase 10.1103/PhysRevB.84.224423}
  {\bibfield  {journal} {\bibinfo  {journal} {Phys. Rev. B}\ }\textbf {\bibinfo
  {volume} {84}},\ \bibinfo {pages} {224423} (\bibinfo {year}
  {2011})}\BibitemShut {NoStop}%
\bibitem [{\citenamefont {Hahn}\ \emph {et~al.}(2014)\citenamefont {Hahn},
  \citenamefont {Naletov}, \citenamefont {de~Loubens}, \citenamefont {Klein},
  \citenamefont {d'Allivy Kelly}, \citenamefont {Anane}, \citenamefont
  {Bernard}, \citenamefont {Jacquet}, \citenamefont {Bortolotti}, \citenamefont
  {Cros}, \citenamefont {Prieto},\ and\ \citenamefont {Mu{\~n}oz}}]{hahn14}%
  \BibitemOpen
  \bibfield  {author} {\bibinfo {author} {\bibfnamefont {C.}~\bibnamefont
  {Hahn}}, \bibinfo {author} {\bibfnamefont {V.~V.}\ \bibnamefont {Naletov}},
  \bibinfo {author} {\bibfnamefont {G.}~\bibnamefont {de~Loubens}}, \bibinfo
  {author} {\bibfnamefont {O.}~\bibnamefont {Klein}}, \bibinfo {author}
  {\bibfnamefont {O.}~\bibnamefont {d'Allivy Kelly}}, \bibinfo {author}
  {\bibfnamefont {A.}~\bibnamefont {Anane}}, \bibinfo {author} {\bibfnamefont
  {R.}~\bibnamefont {Bernard}}, \bibinfo {author} {\bibfnamefont
  {E.}~\bibnamefont {Jacquet}}, \bibinfo {author} {\bibfnamefont
  {P.}~\bibnamefont {Bortolotti}}, \bibinfo {author} {\bibfnamefont
  {V.}~\bibnamefont {Cros}}, \bibinfo {author} {\bibfnamefont {J.~L.}\
  \bibnamefont {Prieto}}, \ and\ \bibinfo {author} {\bibfnamefont
  {M.}~\bibnamefont {Mu{\~n}oz}},\ }\bibfield  {title} {\enquote {\bibinfo
  {title} {Measurement of the intrinsic damping constant in individual
  nanodisks of {Y}$_{3}${Fe}$_{5}${O}$_{12}$ and
  {Y}$_{3}${Fe}$_{5}${O}$_{12}$$|${Pt}},}\ }\href {\doibase 10.1063/1.4871516}
  {\bibfield  {journal} {\bibinfo  {journal} {Appl. Phys. Lett.}\ }\textbf
  {\bibinfo {volume} {104}},\ \bibinfo {pages} {152410} (\bibinfo {year}
  {2014})}\BibitemShut {NoStop}%
\bibitem [{\citenamefont {de~Loubens}\ \emph {et~al.}(2005)\citenamefont
  {de~Loubens}, \citenamefont {Naletov},\ and\ \citenamefont
  {Klein}}]{loubens05}%
  \BibitemOpen
  \bibfield  {author} {\bibinfo {author} {\bibfnamefont {G.}~\bibnamefont
  {de~Loubens}}, \bibinfo {author} {\bibfnamefont {V.~V.}\ \bibnamefont
  {Naletov}}, \ and\ \bibinfo {author} {\bibfnamefont {O.}~\bibnamefont
  {Klein}},\ }\bibfield  {title} {\enquote {\bibinfo {title} {Reduction of the
  spin-wave damping induced by nonlinear effects},}\ }\href {\doibase
  10.1103/PhysRevB.71.180411} {\bibfield  {journal} {\bibinfo  {journal} {Phys.
  Rev. B}\ }\textbf {\bibinfo {volume} {71}},\ \bibinfo {eid} {180411}
  (\bibinfo {year} {2005})}\BibitemShut {NoStop}%
\bibitem [{\citenamefont {Naletov}\ \emph {et~al.}(2003)\citenamefont
  {Naletov}, \citenamefont {Charbois}, \citenamefont {Klein},\ and\
  \citenamefont {Fermon}}]{naletov03}%
  \BibitemOpen
  \bibfield  {author} {\bibinfo {author} {\bibfnamefont {V.~V.}\ \bibnamefont
  {Naletov}}, \bibinfo {author} {\bibfnamefont {V.}~\bibnamefont {Charbois}},
  \bibinfo {author} {\bibfnamefont {O.}~\bibnamefont {Klein}}, \ and\ \bibinfo
  {author} {\bibfnamefont {C.}~\bibnamefont {Fermon}},\ }\bibfield  {title}
  {\enquote {\bibinfo {title} {Quantitative measurement of the ferromagnetic
  resonance signal by force detection},}\ }\href {\doibase 10.1063/1.1614421}
  {\bibfield  {journal} {\bibinfo  {journal} {Appl. Phys. Lett.}\ }\textbf
  {\bibinfo {volume} {83}},\ \bibinfo {pages} {3132} (\bibinfo {year}
  {2003})}\BibitemShut {NoStop}%
\bibitem [{\citenamefont {Capua}\ \emph {et~al.}(2017)\citenamefont {Capua},
  \citenamefont {Rettner}, \citenamefont {Yang}, \citenamefont {Phung},\ and\
  \citenamefont {Parkin}}]{capua17}%
  \BibitemOpen
  \bibfield  {author} {\bibinfo {author} {\bibfnamefont {Amir}\ \bibnamefont
  {Capua}}, \bibinfo {author} {\bibfnamefont {Charles}\ \bibnamefont
  {Rettner}}, \bibinfo {author} {\bibfnamefont {See-Hun}\ \bibnamefont {Yang}},
  \bibinfo {author} {\bibfnamefont {Timothy}\ \bibnamefont {Phung}}, \ and\
  \bibinfo {author} {\bibfnamefont {Stuart S.~P.}\ \bibnamefont {Parkin}},\
  }\bibfield  {title} {\enquote {\bibinfo {title} {Ensemble-averaged rabi
  oscillations in a ferromagnetic {CoFeB} film},}\ }\href {\doibase
  10.1038/ncomms16004} {\bibfield  {journal} {\bibinfo  {journal} {Nature
  Commun.}\ }\textbf {\bibinfo {volume} {8}},\ \bibinfo {pages} {16004}
  (\bibinfo {year} {2017})}\BibitemShut {NoStop}%
\bibitem [{\citenamefont {Bauer}\ \emph {et~al.}(2015)\citenamefont {Bauer},
  \citenamefont {Majchrak}, \citenamefont {Kachel}, \citenamefont {Back},\ and\
  \citenamefont {Woltersdorf}}]{bauer15}%
  \BibitemOpen
  \bibfield  {author} {\bibinfo {author} {\bibfnamefont {Hans~G.}\ \bibnamefont
  {Bauer}}, \bibinfo {author} {\bibfnamefont {Peter}\ \bibnamefont {Majchrak}},
  \bibinfo {author} {\bibfnamefont {Torsten}\ \bibnamefont {Kachel}}, \bibinfo
  {author} {\bibfnamefont {Christian~H.}\ \bibnamefont {Back}}, \ and\ \bibinfo
  {author} {\bibfnamefont {Georg}\ \bibnamefont {Woltersdorf}},\ }\bibfield
  {title} {\enquote {\bibinfo {title} {Nonlinear spin-wave excitations at low
  magnetic bias fields},}\ }\href {\doibase 10.1038/ncomms9274} {\bibfield
  {journal} {\bibinfo  {journal} {Nature Commun.}\ }\textbf {\bibinfo {volume}
  {6}},\ \bibinfo {pages} {8274} (\bibinfo {year} {2015})}\BibitemShut
  {NoStop}%
\bibitem [{\citenamefont {Gerrits}\ \emph {et~al.}(2007)\citenamefont
  {Gerrits}, \citenamefont {Krivosik}, \citenamefont {Schneider}, \citenamefont
  {Patton},\ and\ \citenamefont {Silva}}]{gerrits07}%
  \BibitemOpen
  \bibfield  {author} {\bibinfo {author} {\bibfnamefont {T.}~\bibnamefont
  {Gerrits}}, \bibinfo {author} {\bibfnamefont {P.}~\bibnamefont {Krivosik}},
  \bibinfo {author} {\bibfnamefont {M.~L.}\ \bibnamefont {Schneider}}, \bibinfo
  {author} {\bibfnamefont {C.~E.}\ \bibnamefont {Patton}}, \ and\ \bibinfo
  {author} {\bibfnamefont {T.~J.}\ \bibnamefont {Silva}},\ }\bibfield  {title}
  {\enquote {\bibinfo {title} {Direct {D}etection of {N}onlinear
  {F}erromagnetic {R}esonance in {T}hin {F}ilms by the {M}agneto-{O}ptical
  {K}err {E}ffect},}\ }\href {\doibase 10.1103/PhysRevLett.98.207602}
  {\bibfield  {journal} {\bibinfo  {journal} {Phys. Rev. Lett.}\ }\textbf
  {\bibinfo {volume} {98}},\ \bibinfo {pages} {207602} (\bibinfo {year}
  {2007})}\BibitemShut {NoStop}%
\bibitem [{\citenamefont {L\"uhrmann}\ \emph {et~al.}(1991)\citenamefont
  {L\"uhrmann}, \citenamefont {Ye}, \citenamefont {D\"otsch},\ and\
  \citenamefont {Gerspach}}]{luehrmann91}%
  \BibitemOpen
  \bibfield  {author} {\bibinfo {author} {\bibfnamefont {B.}~\bibnamefont
  {L\"uhrmann}}, \bibinfo {author} {\bibfnamefont {M.}~\bibnamefont {Ye}},
  \bibinfo {author} {\bibfnamefont {H.}~\bibnamefont {D\"otsch}}, \ and\
  \bibinfo {author} {\bibfnamefont {A.}~\bibnamefont {Gerspach}},\ }\bibfield
  {title} {\enquote {\bibinfo {title} {Nonlinearities in the ferrimagnetic
  resonance in epitaxial garnet films},}\ }\href {\doibase
  10.1016/0304-8853(91)90635-N} {\bibfield  {journal} {\bibinfo  {journal} {J.
  Magn. Magn. Mater.}\ }\textbf {\bibinfo {volume} {96}},\ \bibinfo {pages}
  {237--244} (\bibinfo {year} {1991})}\BibitemShut {NoStop}%
\bibitem [{\citenamefont {Watanabe}\ \emph {et~al.}(2017)\citenamefont
  {Watanabe}, \citenamefont {Hirobe}, \citenamefont {Shiomi}, \citenamefont
  {Iguchi}, \citenamefont {Daimon}, \citenamefont {Kameda}, \citenamefont
  {Takahashi},\ and\ \citenamefont {Saitoh}}]{watanabe17}%
  \BibitemOpen
  \bibfield  {author} {\bibinfo {author} {\bibfnamefont {S.}~\bibnamefont
  {Watanabe}}, \bibinfo {author} {\bibfnamefont {D.}~\bibnamefont {Hirobe}},
  \bibinfo {author} {\bibfnamefont {Y.}~\bibnamefont {Shiomi}}, \bibinfo
  {author} {\bibfnamefont {R.}~\bibnamefont {Iguchi}}, \bibinfo {author}
  {\bibfnamefont {S.}~\bibnamefont {Daimon}}, \bibinfo {author} {\bibfnamefont
  {M.}~\bibnamefont {Kameda}}, \bibinfo {author} {\bibfnamefont
  {S.}~\bibnamefont {Takahashi}}, \ and\ \bibinfo {author} {\bibfnamefont
  {E.}~\bibnamefont {Saitoh}},\ }\bibfield  {title} {\enquote {\bibinfo {title}
  {Generation of megahertz-band spin currents using nonlinear spin pumping.}}\
  }\href {\doibase 10.1038/s41598-017-04901-4} {\bibfield  {journal} {\bibinfo
  {journal} {Sci. Rep.}\ }\textbf {\bibinfo {volume} {7}},\ \bibinfo {pages}
  {4576} (\bibinfo {year} {2017})}\BibitemShut {NoStop}%
\bibitem [{\citenamefont {Bonin}\ \emph {et~al.}(2012)\citenamefont {Bonin},
  \citenamefont {d'Aquino}, \citenamefont {Bertotti}, \citenamefont {Serpico},\
  and\ \citenamefont {Mayergoyz}}]{bonin12}%
  \BibitemOpen
  \bibfield  {author} {\bibinfo {author} {\bibfnamefont {R.}~\bibnamefont
  {Bonin}}, \bibinfo {author} {\bibfnamefont {M.}~\bibnamefont {d'Aquino}},
  \bibinfo {author} {\bibfnamefont {G.}~\bibnamefont {Bertotti}}, \bibinfo
  {author} {\bibfnamefont {C.}~\bibnamefont {Serpico}}, \ and\ \bibinfo
  {author} {\bibfnamefont {I.~D.}\ \bibnamefont {Mayergoyz}},\ }\bibfield
  {title} {\enquote {\bibinfo {title} {Analysis of magnetization instability
  patterns in spin-transfer nano-oscillators},}\ }\href {\doibase
  10.1140/epjb/e2011-20505-3} {\bibfield  {journal} {\bibinfo  {journal} {Eur.
  Phys. J. B}\ }\textbf {\bibinfo {volume} {85}},\ \bibinfo {pages} {47}
  (\bibinfo {year} {2012})}\BibitemShut {NoStop}%
\bibitem [{\citenamefont {Thirion}\ \emph {et~al.}(2003)\citenamefont
  {Thirion}, \citenamefont {Wernsdorfer},\ and\ \citenamefont
  {Mailly}}]{thirion03}%
  \BibitemOpen
  \bibfield  {author} {\bibinfo {author} {\bibfnamefont {C.}~\bibnamefont
  {Thirion}}, \bibinfo {author} {\bibfnamefont {W.}~\bibnamefont
  {Wernsdorfer}}, \ and\ \bibinfo {author} {\bibfnamefont {D.}~\bibnamefont
  {Mailly}},\ }\bibfield  {title} {\enquote {\bibinfo {title} {Switching of
  magnetization by nonlinear resonance studied in single nanoparticles},}\
  }\href {\doibase 10.1038/nmat946} {\bibfield  {journal} {\bibinfo  {journal}
  {Nature Mater.}\ }\textbf {\bibinfo {volume} {2}},\ \bibinfo {pages}
  {524--527} (\bibinfo {year} {2003})}\BibitemShut {NoStop}%
\bibitem [{\citenamefont {Pigeau}\ \emph {et~al.}(2011)\citenamefont {Pigeau},
  \citenamefont {de~Loubens}, \citenamefont {Klein}, \citenamefont {Riegler},
  \citenamefont {Lochner}, \citenamefont {Schmidt},\ and\ \citenamefont
  {Molenkamp}}]{pigeau11}%
  \BibitemOpen
  \bibfield  {author} {\bibinfo {author} {\bibfnamefont {B.}~\bibnamefont
  {Pigeau}}, \bibinfo {author} {\bibfnamefont {G.}~\bibnamefont {de~Loubens}},
  \bibinfo {author} {\bibfnamefont {O.}~\bibnamefont {Klein}}, \bibinfo
  {author} {\bibfnamefont {A.}~\bibnamefont {Riegler}}, \bibinfo {author}
  {\bibfnamefont {F.}~\bibnamefont {Lochner}}, \bibinfo {author} {\bibfnamefont
  {G.}~\bibnamefont {Schmidt}}, \ and\ \bibinfo {author} {\bibfnamefont
  {L.~W.}\ \bibnamefont {Molenkamp}},\ }\bibfield  {title} {\enquote {\bibinfo
  {title} {Optimal control of vortex-core polarity by resonant microwave
  pulses},}\ }\href {\doibase 10.1038/nphys1810} {\bibfield  {journal}
  {\bibinfo  {journal} {Nature Phys.}\ }\textbf {\bibinfo {volume} {7}},\
  \bibinfo {pages} {26--31} (\bibinfo {year} {2011})}\BibitemShut {NoStop}%
\bibitem [{\citenamefont {Suto}\ \emph {et~al.}(2018)\citenamefont {Suto},
  \citenamefont {Kanao}, \citenamefont {Nagasawa}, \citenamefont {Mizushima},\
  and\ \citenamefont {Sato}}]{suto18}%
  \BibitemOpen
  \bibfield  {author} {\bibinfo {author} {\bibfnamefont {Hirofumi}\
  \bibnamefont {Suto}}, \bibinfo {author} {\bibfnamefont {Taro}\ \bibnamefont
  {Kanao}}, \bibinfo {author} {\bibfnamefont {Tazumi}\ \bibnamefont
  {Nagasawa}}, \bibinfo {author} {\bibfnamefont {Koichi}\ \bibnamefont
  {Mizushima}}, \ and\ \bibinfo {author} {\bibfnamefont {Rie}\ \bibnamefont
  {Sato}},\ }\bibfield  {title} {\enquote {\bibinfo {title} {Magnetization
  switching of a $\mathrm{Co}/\mathrm{Pt}$ multilayered perpendicular
  nanomagnet assisted by a microwave field with time-varying frequency},}\
  }\href {\doibase 10.1103/PhysRevApplied.9.054011} {\bibfield  {journal}
  {\bibinfo  {journal} {Phys. Rev. Appl.}\ }\textbf {\bibinfo {volume} {9}},\
  \bibinfo {pages} {054011} (\bibinfo {year} {2018})}\BibitemShut {NoStop}%
\bibitem [{\citenamefont {Li}\ \emph {et~al.}(2017)\citenamefont {Li},
  \citenamefont {de~Milly}, \citenamefont {Abreu~Araujo}, \citenamefont
  {Klein}, \citenamefont {Cros}, \citenamefont {Grollier},\ and\ \citenamefont
  {de~Loubens}}]{li17}%
  \BibitemOpen
  \bibfield  {author} {\bibinfo {author} {\bibfnamefont {Yi}~\bibnamefont
  {Li}}, \bibinfo {author} {\bibfnamefont {Xavier}\ \bibnamefont {de~Milly}},
  \bibinfo {author} {\bibfnamefont {Flavio}\ \bibnamefont {Abreu~Araujo}},
  \bibinfo {author} {\bibfnamefont {Olivier}\ \bibnamefont {Klein}}, \bibinfo
  {author} {\bibfnamefont {Vincent}\ \bibnamefont {Cros}}, \bibinfo {author}
  {\bibfnamefont {Julie}\ \bibnamefont {Grollier}}, \ and\ \bibinfo {author}
  {\bibfnamefont {Gr\'egoire}\ \bibnamefont {de~Loubens}},\ }\bibfield  {title}
  {\enquote {\bibinfo {title} {Probing phase coupling between two spin-torque
  nano-oscillators with an external source},}\ }\href {\doibase
  10.1103/PhysRevLett.118.247202} {\bibfield  {journal} {\bibinfo  {journal}
  {Phys. Rev. Lett.}\ }\textbf {\bibinfo {volume} {118}},\ \bibinfo {pages}
  {247202} (\bibinfo {year} {2017})}\BibitemShut {NoStop}%
\bibitem [{\citenamefont {Lavenant}\ \emph {et~al.}(2014)\citenamefont
  {Lavenant}, \citenamefont {Naletov}, \citenamefont {Klein}, \citenamefont
  {de~Loubens}, \citenamefont {Casado},\ and\ \citenamefont
  {Teresa}}]{lavenant14}%
  \BibitemOpen
  \bibfield  {author} {\bibinfo {author} {\bibfnamefont {H.}~\bibnamefont
  {Lavenant}}, \bibinfo {author} {\bibfnamefont {V.}~\bibnamefont {Naletov}},
  \bibinfo {author} {\bibfnamefont {O.}~\bibnamefont {Klein}}, \bibinfo
  {author} {\bibfnamefont {G.}~\bibnamefont {de~Loubens}}, \bibinfo {author}
  {\bibfnamefont {L.}~\bibnamefont {Casado}}, \ and\ \bibinfo {author}
  {\bibfnamefont {J.~M.~De}\ \bibnamefont {Teresa}},\ }\bibfield  {title}
  {\enquote {\bibinfo {title} {Mechanical magnetometry of cobalt nanospheres
  deposited by focused electron beam at the tip of ultra-soft cantilevers},}\
  }\href {\doibase 10.2478/nanofab-2014-0006} {\bibfield  {journal} {\bibinfo
  {journal} {Nanofabrication}\ }\textbf {\bibinfo {volume} {1}},\ \bibinfo
  {pages} {65--73} (\bibinfo {year} {2014})}\BibitemShut {NoStop}%
\bibitem [{mag()}]{magnum}%
  \BibitemOpen
  \href {http://micromagnum.informatik.uni-hamburg.de/} {}\bibinfo {note}
  {{http://micromagnum.informatik.uni-hamburg.de/}}\BibitemShut {NoStop}%
\end{thebibliography}
\end{document}